\documentclass[12pt, a4paper]{article}
\usepackage{jcappub}

\usepackage[utf8]{inputenc}
\usepackage{amsmath, amsthm, amssymb, amsfonts}
\usepackage{bbold}
\pdfoutput=1
\usepackage{float}
\usepackage{booktabs}
\usepackage{graphicx}
\usepackage{subfig}
\usepackage{natbib}
\usepackage{hyperref} 
\usepackage{enumitem}

\usepackage[dvipsnames]{xcolor}
\usepackage{xcolor}
\usepackage{comment}
\usepackage{soul}
\usepackage{url}

\newcommand{\Fermi}{{\sl Fermi}}
\newcommand{\be}{\begin{equation}}
\newcommand{\ee}{\end{equation}}
\newcommand{\bi}{\begin{itemize}}
\newcommand{\ei}{\end{itemize}}
\newcommand{\ben}{\begin{enumerate}}
\newcommand{\een}{\end{enumerate}}

\newcommand{\bx}{{\bf x}}
\newcommand{\btheta}{\boldsymbol{\theta}}
\newcommand{\sigmav}{\langle \sigma v \rangle}
\newcommand{\Msub}{M_{\rm sub}}
\newcommand{\Vmax}{V_{\rm max}}
\newcommand{\LL}{\mathcal{L}}
\newcommand{\mdm}{{M_{\rm DM}}}
\newcommand{\panel}[1]{\textit{#1}}

\usepackage[normalem]{ulem}

\newcommand{\DMsubhaloUnas}{\cite{2010PhRvD..82f3501B, 2012A&A...538A..93Z, 2012ApJ...747..121A, 2012JCAP...11..050Z, 2012PhRvD..86d3504B, 2014GrCo...20...47B, 2014PhRvD..89a6014B, 2015JCAP...12..035B, 2016JCAP...05..028S, 2016ApJ...825...69M, 2017JCAP...04..018H, 2017PhRvD..96f3009C,  2019Galax...7...90C, 2019JCAP...11..045C, 
2019JCAP...07..020C, 2022PhRvD.105h3006C, 2023JCAP...07..033B,  2023MNRAS.520.1348G}}

\newcommand{\DMsubhaloML}{\cite{2016ApJ...825...69M, 2019JCAP...11..045C, 2023JCAP...07..033B, 2023MNRAS.520.1348G}}

\newcommand{\DMsubhaloHand}{\cite{2010PhRvD..82f3501B, 2012A&A...538A..93Z, 2012ApJ...747..121A, 2012JCAP...11..050Z, 2012PhRvD..86d3504B, 2014PhRvD..89a6014B, 2015JCAP...12..035B, 2016JCAP...05..028S, 2017JCAP...04..018H, 2017PhRvD..96f3009C, 2019JCAP...07..020C, 2019JCAP...11..045C}}

\newcommand{\FermiMLclassif}{
\cite{2012ApJ...753...83A, 2016ApJ...825...69M, 2016ApJ...820....8S, Lefaucheur:2017ajt, 2019JCAP...11..045C, 2020MNRAS.492.5377L, 2020MNRAS.495.1093H, 2021RAA....21...15Z, 2021MNRAS.505.5853G, 2022A&A...660A..87B, 2023JCAP...07..033B, 2023MNRAS.521.6195M, 2023RASTI...2..735M, 2023MNRAS.520.1348G}}

\title{Search for dark matter subhalos among unassociated Fermi-LAT sources in presence of dataset shift}

\author[a]{Aurelio Amerio,}
\author[b]{Dmitry V. Malyshev,}
\author[a]{Bryan Zaldívar,}
\author[c]{Viviana Gammaldi,}
\author[d,e]{Miguel A. Sánchez-Conde}

\affiliation[a]{Instituto de F\'isica Corpuscular (IFIC), University of Valencia and CSIC, Calle Catedrático José Beltrán 2, 46980 Paterna, Spain}
\affiliation[b]{Erlangen Centre for Astroparticle Physics,
Nikolaus-Fiebiger-Str. 2, Erlangen 91058, Germany
}
\affiliation[c]{Department of Information Technology, Escuela Politécnica Superior, Universidad San Pablo-CEU, CEU Universities, Campus Montepríncipe, Boadilla del Monte, Madrid 28668, Spain}
\affiliation[d]{Instituto de F\'isica Te\'orica UAM-CSIC,
Universidad Aut\'onoma de Madrid, \\
C/ Nicol\'as Cabrera, 13-15, 28049 Madrid, Spain}
\affiliation[e]{Departamento de F\'isica Te\'orica, M-15,
Universidad Aut\'onoma de Madrid, \\
E-28049 Madrid, Spain}

\emailAdd{aurelio.amerio@ific.uv.es}
\emailAdd{dmitry.malyshev@fau.de}
\emailAdd{b.zaldivar.m@csic.es}
\emailAdd{viviana.gammaldi@ceu.es}
\emailAdd{miguel.sanchezconde@uam.es}

\abstract{We present a search for dark matter (DM) annihilating subhalos of the Milky Way halo among the {\Fermi} Large Area Telescope (LAT) unassociated sources. For this purpose, we construct the first statistical model of the unassociated sources at latitudes above 10 degrees, combining potential DM subhalos with Galactic and extragalactic astrophysical components. 
The distributions of astrophysical sources are constructed based on associated sources, while the DM subhalo distribution is derived from Monte Carlo simulations. We account for differences between associated and unassociated source distributions using a model with covariate and prior probability shifts, which are particular cases of more general dataset shifts. 
This approach is based on quantification learning, advancing beyond previous classify-and-count strategies by providing a well-defined statistical interpretation of the potential contribution from a DM subhalo population. 
For the $b\bar{b}$ annihilation channel and DM masses from 10 GeV to 1 TeV, we find no significant contribution from DM subhalos, and therefore derive 95\% confidence upper limits on the annihilation cross section.
Our analysis yields limits consistent with previous classify-and-count approaches, 
while the underlying generative model provides a more robust statistical framework, opening new avenues for population studies of \Fermi-LAT sources and, more generally, for searches of anomalies, such as a new class of sources in addition to the known classes of sources, in presence of statistical and systematic uncertainties.
}

\begin{document}

\maketitle

\section{Introduction}
\label{sec: intro}
The nature of dark matter (DM) remains one of the biggest mysteries in astrophysics.
N-body cosmological simulations predict that galaxies, such as the Milky Way, are surrounded by a cloud of DM overdensities, which are called subhalos \cite{2001MNRAS.328..726S,2007ApJ...667..859D,2008Natur.454..735D,2009Sci...325..970K,2012AnP...524..507F,2019Galax...7...81Z,2020NatRP...2...42V}.
Subhalos with masses above ${\sim 10^8~M_{\odot}}$ are expected to contain stars: these subhalos are detected as dwarf spheroidal galaxies.
Smaller and more numerous subhalos may have no usual matter (such as stars or dust) inside them.
In this paper, we focus on subhalos with masses smaller than ${\sim 10^8~M_{\odot}}$, which are not associated to
dwarf spheroidal galaxies.
These subhalos can be detected either by gravitational effects or through annihilation or decay of DM particles into Standard Model (SM) particles.
One of the most popular types of particle DM candidates is the Weakly Interacting Massive Particle (WIMP) with masses in the GeV -- TeV range annihilating into SM particles, such as $b\bar{b}$ or $\tau^+\tau^-$ 
\cite{2016PhR...636....1C, 2017NatPh..13..224C, MiniReview_Gammaldi}.
For sufficiently large masses of DM particles, e.g., above about 1 GeV, DM annihilation events in subhalos can be detected by the {\Fermi} Large Area Telescope (LAT).
Although DM subhalos may emit radiation at other frequencies, e.g., in radio wavelengths due to synchrotron radiation from the electrons and positrons created in DM annihilations, the corresponding signal is expected to be small for DM subhalos with masses less than ${\sim 10^8~M_{\odot}}$ \cite{2015JCAP...02..032C} (the radio emission from the dwarf spheroidal galaxies can be rather significant but has large uncertainties \cite{2007PhRvD..75b3513C, 2013PhRvD..88h3535N, 2014JCAP...10..016R}).
As a result, the DM subhalos would primarily manifest themselves as a population of unassociated gamma-ray sources in {\Fermi}-LAT catalogs with approximately isotropic distribution in the sky 
\DMsubhaloUnas.

The latest \Fermi-LAT catalog (4FGL-DR4)~\cite{2022ApJS..260...53A, 2023arXiv230712546B} contains 2428 unassociated sources (unIDs), which is about one third of all detected sources.
Among these sources, 1282 are at latitudes $|b| > 10^\circ$.
Provided that unIDs have no known counterparts at other wavelengths, these gamma-ray sources are natural candidates to search for DM-dominated subhalos shining in gamma rays
due to DM annihilation.
The problem is that the majority of unIDs are expected to be astrophysical sources, where the lack of association can be due to poor localization of the gamma-ray source or, in case of pulsars, beaming effects.

The search for DM subhalos has been previously performed with the ``classify-and-count'' strategy, where a number of DM subhalo candidates has been selected either based on their gamma-ray properties 
consistent with the expectations for DM subhalos~\DMsubhaloHand, or based on classification of gamma-ray sources with machine learning (ML)~\DMsubhaloML.
The limits on DM annihilation have been derived assuming that there cannot be more DM subhalos among \Fermi-LAT unIDs than the number of DM subhalo candidates.

A general caveat of previous searches of DM candidates based on gamma-ray properties is that the same source may be consistent both with a DM subhalo as well as an astrophysical source, e.g., an active galactic nucleus (AGN) or a pulsar.
Thus, unless there is a smoking gun signature of a DM subhalo, such as a physical extension consistent with the distribution of DM 
\cite{2014PhRvD..89a6014B, 2015JCAP...12..035B, 2017PhRvD..96f3009C, 2022PhRvD.105h3006C, 2019JCAP...11..045C}, this analysis does not allow one to claim a detection of a population of DM subhalos among \Fermi-LAT unIDs.
Although one can estimate an upper limit on DM annihilation assuming that the number of DM subhalos cannot exceed the number of DM subhalo candidates,
it is, nevertheless, challenging to calculate a statistical significance of the DM exclusion.
For instance, in ML-based approaches the estimate of the number of DM subhalo candidates depends on the threshold for the probability that an unassociated source is a DM subhalo.
The threshold is selected by hand, i.e., it is not based on a statistical argument. In addition, the probability itself that a source is a DM subhalo depends on the training dataset, e.g., on the choice of the fractions of DM-like and astrophysical sources. Usually one uses 50/50\% split of astrophysical and DM subhalo sources based on the requirement that the classes are balanced in the training dataset (smaller or larger fractions of DM-like sources in the training dataset would result in respectively smaller or larger predicted DM-like probabilities). However, the 50\% fraction of DM-like sources in the training dataset does not come from data, provided that there are no DM subhalos among associated sources. If one, for example, uses balanced classes for classification of sources into pulsars and AGNs, then one would predict a larger fraction of pulsars among unassociated sources
than what would be expected from associated sources, where there are almost 10 times more AGNs than pulsars. Such bias coming from balancing of classes has to be corrected (calibrated) in order to obtain realistic class probabilities.
For the calibration one would use a realistic testing dataset, which is not available for the DM subhalos (since none of them has been found in gamma-ray data).
The problem of difference of class prevalences for training and testing (or target) datasets is known as prior shift.
Similar problem (implicitly) exist for searches of DM subhalos, which are not based on ML estimates of DM-like probabilities.

Another problem, which affects the searches of DM subhalos but has not been addressed in the literature so far (e.g., in the classify-and-count approach) is that the distribution of associated and unassociated sources are different as a function of most of features, such as flux, spectral index, spectral curvature, variability etc.~\cite{2022ApJS..260...53A, 2023RASTI...2..735M}.
This difference can be due to bias in associations, e.g., it is easier to associate bright sources at high latitudes (cf. covariate shift discussion in section~\ref{sec:cov_prior}). It may also be a signature of a new class of sources, such as DM subhalos (cf. prior shift discussion above and in section~\ref{sec:cov_prior}).
A priori the reason for the differences in the distributions is not known. As a result, one needs to put additional constraints, e.g., on the association bias, in order to break the degeneracy between the association bias and the presence of a new class of sources.

In other words, the distributions of associated and unassociated \Fermi-LAT gamma-ray sources are different as a function of source parameters.
This difference may be due to association bias (described by covariate shift), different fractions of astrophysical classes of sources (described by prior shift), or there may be a new class of sources, e.g., DM subhalos (which is a particular case of a prior shift - from zero to non-zero prevalence).
A solution to this problem is not possible in the classify-and-count approach, since it relies on the prior assumptions of the prevalence of classes, which cannot be adjusted in the absence of a realistic testing set.
In order to overcome the limitation of the classify-and-count approach we use, for the first time in searches of DM subhalos, a different analysis method referred as quantification learning rather than classification.

The main purpose of quantification analysis 
\cite{10.1145/3117807_Gonzalez_quantification, 2022arXiv221108063M, esuli2023learning} 
is to accurately predict the prevalence of classes rather than to optimize the performance for classification of individual sources, which is the main goal of classification analysis.
One of the methods in quantification learning is known as template matching~\cite{2024arXiv240100490M}. It is familiar in astrophysics in modeling of large-scale radio, X-ray, or gamma-ray emission as a linear combination of templates based on tracers of the different components of emission, e.g., the distribution of dust is used as a tracer of hadronic component of diffuse gamma-ray emission due to interactions of high-energy cosmic rays with gas.

In this paper, we construct for the first time a statistical model for the distribution of unassociated sources at latitudes $|b| > 10^\circ$.%
\footnote{We exclude sources with $|b| < 10^\circ$, since the main background in this analysis is the astrophysical sources among the unassociated ones. Almost half of the unassociated sources (1146 out of 2428) are within this range of latitudes and most of them are expected to be the astrophysical Galactic sources. The decrease in the area, on the other hand, is about 17\%, i.e., assuming an isotropic distribution of DM subhalos around the Earth, the increase in signal-to-noise ratio due to masking of the Galactic plane can be estimated as $ (1. - 0.17) \sqrt{2428/1146} \approx 1.14$.
This increase is likely larger due to higher flux detection threshold (smaller detectability volume for DM subhalos) toward the Galactic plane, i.e., the decrease in signal is less than 17\% as a result of masking the Galactic plane.
}
It consists of a mixture model, i.e. a linear combination of probability distribution functions (PDFs)
of classes of astrophysical sources plus a contribution of DM subhalos.
The PDFs of classes of astrophysical sources are determined from the associated sources, while the distribution of DM subhalos is determined from Monte Carlo simulations assuming a model of J-factors from N-body DM simulations and particular parameters for DM (mass, annihilation cross section, and channel).
The prior shift problem is addressed by fitting the relative contribution of the different components to the data (the distribution of unassociated sources), while the covariate shift problem is addressed by adding a modulation to the PDFs of the classes of associated sources (the parameters of the modulation are also fit to the data simultaneously with the class prevalence).

In more formal terms, 
we construct PDFs of astrophysical classes, denoted as $p(\bx|k)$,
which represent the probability of a source with features $\bx$ given a source class $k$. 
The PDFs for unassociated sources are determined by modulating the PDFs of the associated ones by functions of input features $\bx$ (these functions are the same for all classes $k$).
Previous supervised learning studies have focused on estimating the conditional probability $p(k|\bx)$ of the class of source $k$ given the input features 
\FermiMLclassif, as typically done by probabilistic classification models.
Estimating $p(\bx|k)$ instead brings several crucial advantages. First of all (see section~\ref{sec:approach}), it allows us to model the difference between distributions of associated and unassociated sources as a combination of both covariate and prior shifts.
The latter allows us to model the different prevalence of classes of astrophysical sources for training (associated sources) and target (unassociated sources) datasets as well as to include a new class of sources among the unassociated ones, which in this analysis is modeled as DM subhalos.
Moreover, upon specification of the class prior $p(k)$, the joint distribution $p(k,\bx)=p(\bx|k)p(k)$ can be sampled to generate mock data, which can be used in Monte Carlo analyses.\footnote{That's why probabilistic models estimating $p(\bx|k)$ are called ``generative models'', as opposed to ``discriminative'' models focusing on $p(k|\bx)$.} Finally, the joint PDF  $p(\bx, k)$ can be marginalized over the class $k$ to estimate $p(\bx)$, i.e., the marginal distribution of the source features. 
The quantity $p(\bx)$
represents a probability density for a source with parameters $\bx$. 
The product of $p(\bx_i)$ over unassociated sources is the model likelihood, which we maximize to obtain the optimum model parameters both for the prior and covariate shifts.
In particular, we use such likelihood to estimate the best fit and upper bounds on the DM annihilation cross section $\sigmav$.
Although generative models have been constructed previously in analyses of \Fermi-LAT source distributions, e.g. \cite{2023MNRAS.520.1348G}, they were not used for construction of maximum likelihood models, which we do in the current work.

The paper is organized as follows.
In section~\ref{sec:data},
we describe the selection of the data.
We discuss the covariate and prior shifts in section~\ref{sec:cov_prior}. We present our statistical model in section~\ref{sec:stat-model}. 
The model of J-factors used in this paper is reviewed in section~\ref{sec:sim-jfactor}, while the simulation of gamma-ray sources corresponding to DM subhalos is presented in section~\ref{sec:DM_model}. 
We optimize the model in section~\ref{sec:model-opt}, while
the corresponding limits on DM annihilation determined from the observed gamma-ray sources are presented in section~\ref{sec:dm-limits}.
Section~\ref{sec:conclusions} contains discussion and conclusions. 
We provide details about the statistical model and
perform several consistency tests in appendices.

\section{Statistical analysis}
\label{sec:approach}

\subsection{Data selection}
\label{sec:data}

In this work we use the 4FGL-DR4 catalog~\cite{2022ApJS..260...53A, 2023arXiv230712546B}. We group the known types of associated sources into two classes:%
\footnote{Cf. table 5 in ref.~\cite{2022ApJS..260...53A} for details about the types of sources. Galactic sources: psr, hmb, sfr, snr, pwn, gc, gal, bin, msp, lmb, spp, glc, nov. Extragalactic sources: bll, sbg, rdg, css, bcu, ssrq, fsrq, sey, nlsy1, agn.}
\begin{itemize}
    \item Galactic {(dominated by millisecond pulsars at high latitudes~\cite{smith2023third, 2020ApJS..247...33A, Fornasa:2015qua, Hooper:2024avz, Cholis:2014noa, Caraveo:2013lra})}: pulsars, binaries, star-forming regions, supernova remnants, pulsar wind nebulae, globular clusters, and novae;
    \item Extragalactic ({mostly AGNs~\cite{fossati1998unifying, 2020ApJS..247...33A, Fornasa:2015qua, Dermer:2009zz, Fermi-LAT:2014ryh, Ajello:2015mfa, Lisanti:2016jub}}): blazars, quasars, starburst galaxies, normal galaxies, radio galaxies and radio sources, Seyfert galaxies, and non-blazar active galaxies.
\end{itemize}
{This choice is made to group sources based on their fundamentally different physical origins and radiation environments. By separating the magnetospheric emission characteristic of Galactic pulsars from the jet-driven processes of extragalactic AGNs, we ensure that the classification reflects the intrinsic physical diversity of the $\gamma$-ray sky, which manifests in distinct spectral signatures.}

\begin{figure}
    \centering
    \includegraphics[width=0.45\linewidth]{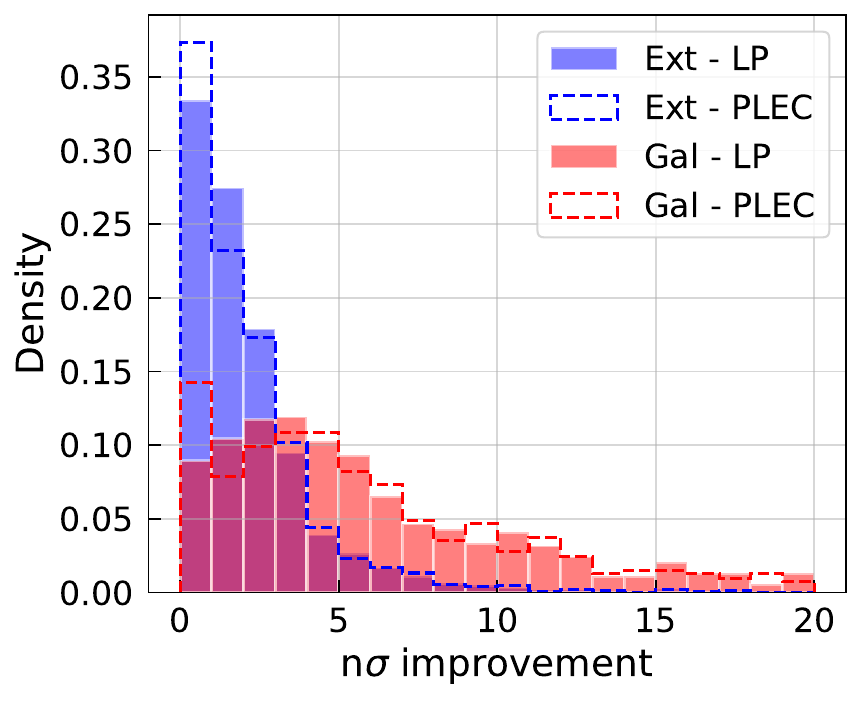}
    \includegraphics[width=0.45\linewidth]{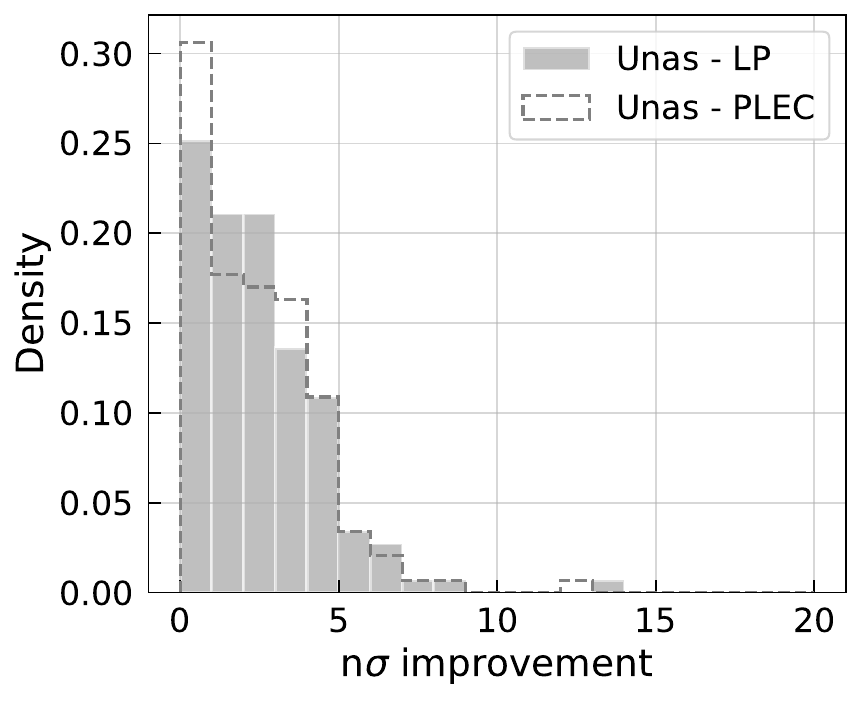}
    \caption{{Distribution of the significance of fit improvement, as reported in the 4FGL-DR4 catalog, when replacing a power-law energy spectrum with either a log parabola (LP) or a power law with an exponential cutoff (PLEC). We display the histograms for associated sources ({left} panel) and unassociated sources ({right} panel).}}
    \label{fig:LP_vs_PLEC}
\end{figure}

In order to characterise the sources, we focus on their energy spectra modeled with the log-parabola (LP) function
\be
\label{eq:log_par}
\ln \frac{F(E)}{K} = - \alpha \ln \frac{E}{E_0} - \beta \left(\ln \frac{E}{E_0} \right)^2,
\ee
where the parameters correspond to the following columns in the 4FGL-DR4 catalog:
$K$ -- {\tt LP\_ Flux\_Density},
$\alpha$ -- {\tt LP\_Index},
$\beta$ -- {\tt LP\_beta},
and $E_0$ -- {\tt Pivot\_Energy}.
\\
{The LP model provides a well-motivated spectral choice; it offers a significant improvement over a simple power law, as evidenced in the \texttt{LP\_SigCurv} column in the 4FGL-DR4 catalog, and performs at least as well as a power law with exponential cut off both for the unassociated and for the associated Galactic and extragalactic sources (figure~\ref{fig:LP_vs_PLEC}).}

Provided that only three parameters are independent and that $E_0$ in the catalog is different for different sources, we rescale $E_0$ to a fixed value of 1 GeV (for analysis of $\mdm> 20$ GeV) or 100 MeV (for $\mdm < 20$ GeV). 
We use the following three parameters as input features:

\ben
\item 
$\log_{10}\phi$ -- the logarithm of the
 flux integrated between 100 MeV and 1 TeV (obtained by summing fluxes in energy bins provided in the catalog);
\item 
$\alpha$ -- {\tt LP\_Index} rescaled to $E_0 = 1$ GeV or $E_0 = 100$ MeV depending on the choice of DM mass, i.e., above or below 20 GeV respectively;
\item
$\beta$ -- {\tt LP\_beta}.
\een
We collect these variables for each source $i$ inside a vector ${\bx}_i\equiv \{\log_{10}(\phi_i), \alpha_i, \beta_i\}$.
In our analysis, we consider only sources with Galactic latitudes $|b| > 10^\circ$.

\subsection{Covariate and prior shifts}
\label{sec:cov_prior}

\begin{figure}
    \centering
    \includegraphics[width=1.0\textwidth]{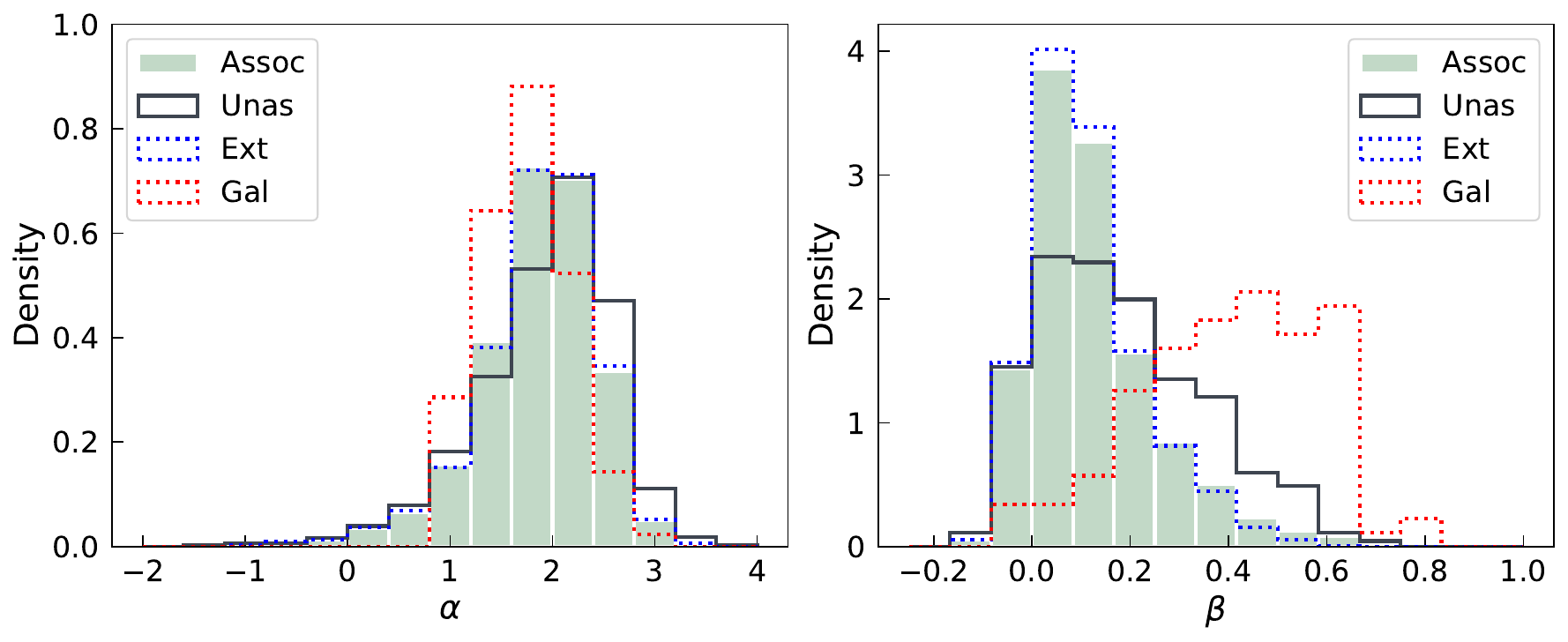}
    \caption{Normalised distributions of $\alpha$ and $\beta$ parameters for associated and unassociated sources for $E_0=1$ GeV, as well as the distributions for extragalactic and Galactic sources with $|b| > 10^\circ$.}
    \label{fig:alpha_beta_hist_E01000MeV}
\end{figure}

The basic assumption behind a supervised classification task is that the joint distribution of the input features $\bx$
and output classes (type of sources) $k$  are the same for the training and target datasets:
\be
p_{\rm train} (\bx, k) = p_{\rm target} (\bx, k),
\ee
while, in general, a dataset shift represents a situation where the training and target distributions are different
$p_{\rm train} (\bx, k) \neq p_{\rm target} (\bx, k)$.
In our case, the training dataset is the associated sources, while the target dataset is the unassociated sources.
Consequently, we will use associated (assoc) and unassociated (unas) subscripts for the training and target datasets, respectively.
\\
The joint distribution of features $\bx$ and class labels $k$ can be expressed as a product of conditional and marginal probabilities in two equivalent ways:
\be
p(\bx, k) = p(k|\bx) p(\bx) = p(\bx|k) p(k).
\ee
This decomposition gives rise to two distinct types of dataset shift~\citep{MorenoTorres2012AUV}:
\ben
\item
{ Covariate shift}: the conditional class probabilities remain unchanged,\\
{$p_{\rm assoc}(k|\bx) = p_{\rm unas}(k|\bx)$}, but the feature distributions differ, $p_{\rm assoc}(\bx) \neq p_{\rm unas}(\bx)$.
\item
{ Prior shift}: the feature distributions conditioned on the class are invariant, $p_{\rm assoc}(\bx|k) = p_{\rm unas}(\bx|k)$, but the class prevalence changes, $p_{\rm assoc}(k) \neq p_{\rm unas}(k)$.
\een
Note that for the covariate shift, differences in the marginal distribution of the features $p(\bx)$ generally lead to differences in the marginal class probabilities $p(k)$.
Analogously, for prior shift, changes in the marginal class distributions $p(k)$ in general result in changes in the marginal feature distributions $p(\bx)$. We show the distributions of the $\alpha$ and $\beta$ parameters in figure~\ref{fig:alpha_beta_hist_E01000MeV}. Notice that the distributions are different for associated and unassociated sources.
In principle, this difference can be due to either a covariate shift or a prior shift. 
For example, as can be seen from the histograms of the associated extragalactic and Galactic sources, a change in the class prevalence in unassociated sources would certainly lead to a change in the $\beta$ parameter distribution, which could be explained as a prior shift.
On the other hand, if there are differences in the distribution of unassociated sources, which affect all unassociated sources, e.g., larger uncertainties in the reconstructed energy spectrum, 
then this effect can be addressed by a covariate shift.

\subsection{Statistical model}
\label{sec:stat-model}

In general, our goal is to construct a probabilistic model for the distribution of unassociated sources $\tilde{p}_{\rm unas}(\bx)$ under the assumption that it is a mixture of classes of associated sources plus a possible new class of sources.

In the covariate shift assumption (in absence of a prior shift) this task is actually trivial as
\be
\label{eq:cov0}
p_{\rm unas}(\bx) 
= \sum_k p_{\rm unas}(k|\bx) p_{\rm unas}(\bx) 
= \sum_k p_{\rm assoc}(k|\bx) p_{\rm assoc}(\bx) C(\bx),
\ee
where in the second equality we use the covariate shift assumption $p_{\rm assoc}(k|\bx) = p_{\rm unas}(k|\bx)$ and define $C(\bx) = p_{\rm unas}(\bx) / p_{\rm assoc}(\bx)$.
The function $C(\bx)$ describes the covariate shift, i.e., the difference in the distribution of training and target datasets.

\begin{figure}
    \centering
    \includegraphics[width=0.95\textwidth]{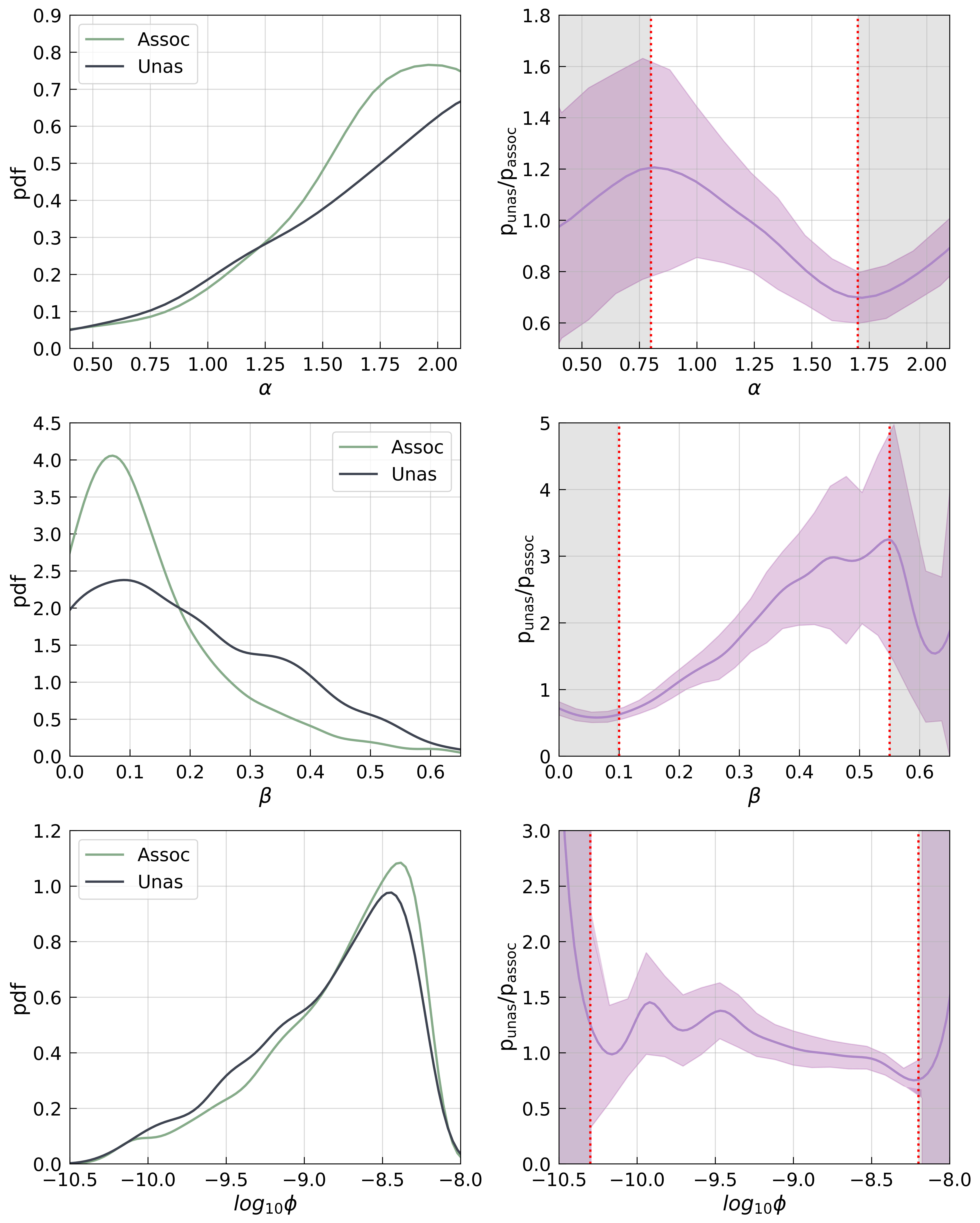}
    \caption{\textit{Left column:} full distributions of associated and unassociated sources modeled with KDE for the case of $E_0$ = 1 GeV before truncation and normalization to the ROI. 
    \textit{Right column:} the ratio of the distributions $p_{\rm unas}/p_{\rm assoc}$. The red vertical lines show the range of parameters that we use in the analysis (ROI). We estimate the error (purple bands) as the Poisson error in the histograms and compute the relative errors for the associated and unassociated distributions, which we sum in quadrature. 
    }
    \label{fig:dist_ratio}
\end{figure}

Moreover, we can add a new class of sources with a PDF 
$p_{\rm new}(\bx)$ and define a model for the unassociated sources as:
\be
\label{eq:cov1}
\tilde{p}_{\rm unas}(\bx) 
= \sum_k p_{\rm assoc}(k|\bx) p_{\rm assoc}(\bx) \tilde{C}(\bx) + \pi_{\rm new} p_{\rm new}(\bx),
\ee
where 
$\tilde{C}(\bx) = {\rm max}(p_{\rm unas}(\bx) - \pi_{\rm new}p_{\rm new}(\bx), 0) / p_{\rm assoc}(\bx)$ can be viewed as a generalized covariate shift that describes the difference in the distribution of training dataset and a part of the target dataset that corresponds to the same classes as in the training dataset.
If $p_{\rm unas}(\bx) \geq \pi_{\rm new}p_{\rm new}(\bx)$ for $\forall \bx$, then the model in eq.~(\ref{eq:cov0}) is equal to the model in eq.~(\ref{eq:cov1}), i.e., the two models are indistinguishable until the new class predicts more sources in some region of space than there are unassociated sources.
Although mathematically we cannot exclude a situation when all unassociated sources in some region of space belong to a new class, this is not very likely from the physical point of view.
In order for the model to be more constraining we put additional physical conditions on the generalized covariate shift function $\tilde{C}(\bx)$.
In particular, we model $\tilde{C}(\bx)$ as a product of 1-dimentional functions
\be
\label{eq:cov_model}
\tilde{C}(\bx) = \prod_{d=1}^3 f_d(x_d),
\ee
where each $f_d(x_d)$ is monotonous. 
The product is over the dimensions $d$ of the space of input variables. 
In particular, we assume here that the covariate shift in each of the input variables is independent of the other input variables.  
We implement $f_d(x_d)$ using sigmoid functions
\be
\label{eq:sigmoid}
f(x_d;b_d,c_d) = \frac{1}
{1+e^{-b_d(x_d - c_d)}},
\ee
where each function has two parameters $(b_d,c_d)$ determined from maximizing the likelihood of the model. 
Our region of interest (ROI) in feature space will be restricted to a window in every dimension where the ratio 
$p_{\rm unas}/p_{\rm assoc}$ 
of the empirical distributions are monotonous. 
We approximate the empirical distributions of associated and unassociated gamma-ray sources using kernel density estimation (KDE) with a Gaussian kernel \cite{bishop}. 
In order to find the optimal bandwidth for the kernels, we perform 5-fold cross-validation repeated 20 times. 
For each of the repetitions, we randomize the dataset,  split it into 5 subsets, and repeat the training 5 times: each time, a new subset is used for testing, and the remaining four subsets are used for training.
The performance is evaluated by calculating the likelihood of the model on the testing dataset.
The optimal bandwidth parameter is obtained from the maximum of the performance averaged over the 100 evaluations (20 repetitions with 5 training/testing splits in each repetition).
The optimal bandwidth parameters for associated and unassociated sources are computed independently.
We show the corresponding KDE models in the $E_0 = 1$ GeV case in the left panels in figure~\ref{fig:dist_ratio}.
The right panels of figure~\ref{fig:dist_ratio} show the ratios of the KDE models, where the bands represent the statistical uncertainty.

The ROI is determined as the intervals in parameter space where the ratio of the KDE models is monotonous within statistical uncertainties.
In the $E_0 = 1$ GeV case, we restrict our analysis to $\alpha \in [0.8,\, 1.7]$.
Analogously, in the $E_0 = 100$ MeV case, $\alpha \in [0.2,\, 0.8]$.
The range of the $\beta$ parameter is the same in both cases, $[0.1, \, 0.55]$.
Finally, we restrict the $\log_{10}\phi$ parameter to the interval $[-10.3, \, -8.2]$.
In addition, we compute the 3D KDE models and identify the optimal bandwidth parameters independently for Galactic and extragalactic classes of sources, which yields the $p_{\rm assoc}(\bx|k)$ used in the analysis.

In general, the weight (prevalence) of classes of sources does not need to be the same for associated and unassociated sources, i.e., there can be a prior probability shift in addition to the covariate shift.
In order to allow a change in the prevalence of the classes of sources, we express the terms in the sum in eq.~(\ref{eq:cov1}) in the form suitable for the prior shift:
\be
\label{eq:p_assoc}
p_{\rm assoc}(k|\bx) p_{\rm assoc}(\bx) \equiv
p_{\rm assoc}(\bx, k) = \pi_k p_{\rm assoc}(\bx|k),
\ee
where for associated sources $\pi_k = \pi_k^{\rm assoc}$ is the prevalence of class $k$, while for unassociated sources, $\pi_k$ are in general unknown and will be determined from the fit of the model to the data.
Overall, our model of the unassociated source takes the form:
\be
\label{eq:mixture}
\tilde{p}_{\rm unas}(\bx) 
= \left(\sum_{k=1}^K \pi_k p_{\rm assoc}(\bx|k)\right) \tilde{C}(\bx; \btheta_{\rm cov}) + \pi_{\rm DM}(\btheta_{\rm DM}) p_{\rm DM}(\bx; \btheta_{\rm DM}),
\ee
where $K = 2$ is the number of classes of associated sources (Galactic and extragalactic, cf. section~\ref{sec:data}). 
The DM distribution $p_{\rm DM}(\bx; \btheta_{\rm DM})$ is determined below in section~\ref{sec:DM_model}. 

The optimizable parameters in this model are: $\pi_k$, describing the prevalence of the classes of associated sources; $\btheta_{\rm cov}\equiv\{b_d,c_d\}$ for $d = 1,\,2,\,3$, parametrizing the covariate shift in eq.~(\ref{eq:sigmoid}), and $\sigmav$, the annihilation cross section, which we have included in the definition of $\btheta_{\rm DM}=\{\sigmav,\mdm,{\rm ch.}\}$. The other parameters, namely the DM mass $\mdm$ as well as the annihilation channel `ch.', are kept fixed during the optimization procedure.

Note that for $\tilde{C}(\bx; \btheta_{\rm cov}) = 1$, the model reduces to a prior shift model, while for $\pi_k = \pi_k^{\rm assoc}$ and $\pi_{\rm DM} = 0$ the model reduces to a covariate shift model.
Thus, in general, the model in eq.~(\ref{eq:mixture}) is a combination of covariate and prior shift models. We point out that the model in eq.~(\ref{eq:mixture}), identified as the likelihood of a given unassociated source, allows us to:
\ben
\item
determine the unknown parameters of the model by maximizing the likelihood;
\item
detect or put an upper bound at a given level of confidence for a possible new component, such as DM;
\item
generate new data (for Monte Carlo tests);
\item
determine the conditional class probabilities of individual sources using the probability rule
$p(k|\bx) = p(\bx, k)/\tilde{p}_{\rm unas}(\bx)$. In the case of astrophysical classes 
$p(\bx, k) = \pi_k p_{\rm assoc}(\bx|k) \tilde{C}(\bx; \btheta_{\rm cov})$, while
for the DM class: $p(\bx, {\rm DM}) = \pi_{\rm DM} p_{\rm DM}(\bx; \btheta_{\rm DM})$.
\een

We stress that, as far as we are aware of, this is the first time a generative probabilistic model for the unassociated sources has been proposed. Previous studies have focused on modeling directly the conditional probabilities $p(k|\bx)$, which does not give access to the data likelihood. In this respect, our approach is more informative.

\section{Dark matter subhalos model}
\label{sec:DM_model_chapter}

\subsection{Subhalo J-factors from N-body simulations}
\label{sec:sim-jfactor}
A key component of our pipeline is the so-called ‘J-factor distribution’ of the Galactic subhalo population. This J-factor, also known simply as ‘astrophysical factor’, acts as a normalization factor for the expected annihilation signal and can be computed as the volume integral of the subhalo DM density squared. In this work, we follow ref.~\cite{2024MNRAS.530.2496A} and adopt the J-factor values provided for two different prescriptions of the subhalo internal structure%
\footnote{All data are available at \url{https://projects.ift.uam-csic.es/damasco/?page_id=831}.}: one derived from the mass enclosed within the subhalo tidal radius, $J_{\rm tidal}^{\Msub}$, and another based on the maximum circular velocity of each subhalo, $J_{\rm tidal}^{\Vmax}$. 
These two prescriptions are advisable here to estimate current J-factor uncertainties. 
Indeed, measuring subhalo masses is not a trivial task, especially at scales close to the numerical resolution limit of the simulation, where only a few simulation particles are available. Because of this, the computation of masses relies on adopting a specific form of the subhalo DM density profile, which is then fitted to the available simulation data and integrated up to the assumed tidal radius. Unfortunately, even the precise value of the latter can vary significantly among different definitions and is very sensitive to changes in the tidal forces acting on the subhalo as it orbits around the host halo center. In contrast to these mass-related issues, the peak of the circular velocity profile within each subhalo, i.e. $V_{max}$, has been shown to be less affected by numerical resolution and tidal forces (see, e.g., refs.~\cite{2024MNRAS.530.2496A,2025arXiv250601152A}). Its determination also does not depend on adopting a specific functional form of the DM density profile to model the subhalo structural properties. As a result, J-factor calculations based on the latter approach are considered more reliable. Nevertheless, we use both J-factor sets in this study mainly to enable direct comparison with previous results, where J-factor based on subhalo masses are more common. Also, as said, the exercise will  be useful to understand the current level of uncertainties on subhalo modeling.

The calculation of astrophysical J-factors in ref.~\cite{2024MNRAS.530.2496A} is based on the Via Lactea II (VL-II) N-body cosmological simulation \cite{2007ApJ...667..859D}, a DM-only simulation of a Milky-Way-sized halo resolving subhalos down to approximately $10^4$ ${\rm M}_{\odot}$. 
To explore subhalos below the VL-II resolution limit, 
the simulation was repopulated with millions of smaller subhalos, extending the analysis down by nearly five orders of magnitude 
to about $10^{-1}$ ${\rm M}_{\odot}$~\cite{2022PhRvD.105h3006C}.
Additionally, by placing the Earth at random positions consistent with its Galactocentric distance, 
thousands of realizations of a Milky Way-like galaxy were created to obtain statistically robust results. 
This led to accurate J-factor distributions (based on both mass and circular velocity) for the entire subhalo population of a galaxy like our own in a $\Lambda$CDM universe. A limitation of ref.~\cite{2024MNRAS.530.2496A} is the absence of baryonic effects in the parent simulation. Baryons are expected to significantly influence the DM subhalo population, particularly in the Galaxy's innermost regions (e.g.,~\cite{2019MNRAS.487.4409K,2017MNRAS.471.1709G,2021MNRAS.501.3558G}), however none of these hydrodynamical simulations resolve subhalos lighter than $\sim 10^6$ ${\rm M}_{\odot}$. Thus, the survival of such low-mass subhalos to tidal stripping, especially with baryonic influences, remains uncertain~\cite{2018MNRAS.474.3043V,2020MNRAS.491.4591E,2023MNRAS.518...93A}.

In addition to $J_{\rm tidal}^{\Msub}$ and $J_{\rm tidal}^{\Vmax}$ from ref.~\cite{2024MNRAS.530.2496A}, we also utilize $J_{0.1}^{\Msub}$ and $J_{0.1}^{\Vmax}$, derived from the same simulation work. These values correspond to J-factors integrated not up to the tidal radius of each subhalo but rather to the corresponding radius subtending 0.1$^\circ$ in the sky. 
This choice aligns with the typical angular resolution of the {\Fermi}-LAT in the GeV range. By using $J_{0.1}^{\Msub}$ and $J_{0.1}^{\Vmax}$, we focus on point-like subhalo emissions, as would appear in \Fermi-LAT point-source catalogs. 
Notably, ref.~\cite{2022PhRvD.105h3006C} concludes that the brightest Galactic subhalos in the \Fermi-LAT gamma-ray sky, if detected, would be extended, with angular sizes around $0.2^\circ$ – $0.3^\circ$. 
Consequently, for such subhalos, our $J_{0.1}^{\Msub}$ and $J_{0.1}^{\Vmax}$ values are expected to be significantly lower than their corresponding $J_{\rm tidal}^{\Msub}$ and $J_{\rm tidal}^{\Vmax}$ values. 
We adopt $J_{0.1}^{\Vmax}$ as our nominal case in this paper, while we use the other J-factor values to compare with previous literature, e.g., in Calore et al. (2019)~\cite{2019Galax...7...90C} a $J_{0.1}^{M_{\rm sub}}$ model was used.

\begin{figure}
    \centering
    \includegraphics[width=0.5\linewidth]{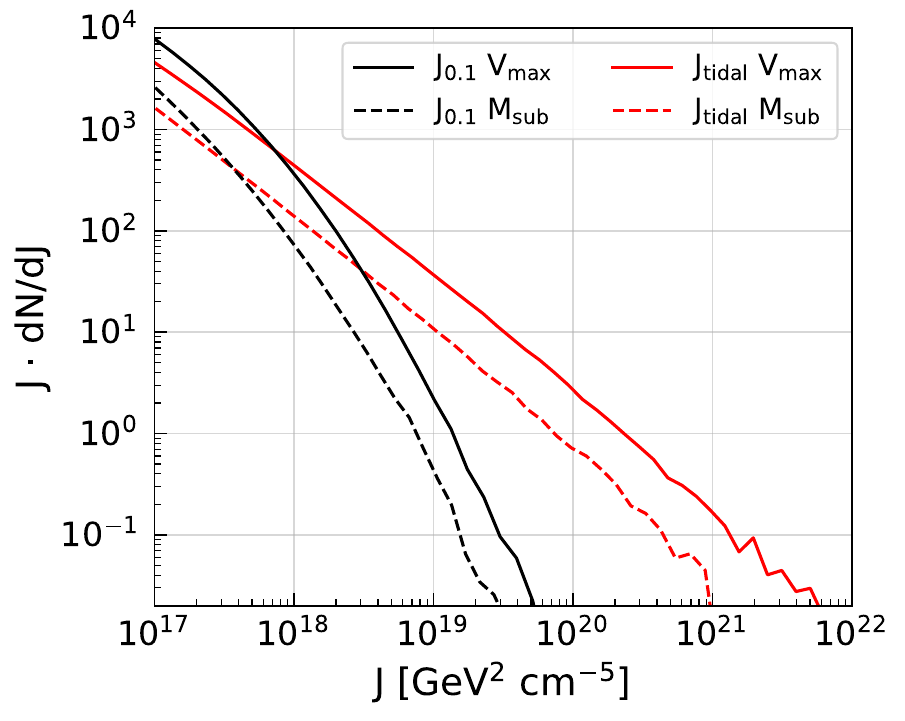}
    \caption{J-factor number density times J for the different J-factor cases considered in this work. See section~\ref{sec:sim-jfactor} for further details.}
    \label{fig:Jfactors}
\end{figure}

Figure~\ref{fig:Jfactors} illustrates the four sets of J-factor distributions discussed in this section, along with their corresponding number density, which will be the relevant quantity for our pipeline.

\subsection{Gamma-ray emission from dark matter subhalos}
\label{sec:DM_model}

Given a DM setup $\btheta_{\rm DM}$ (i.e., a DM mass ${\mdm}$, annihilation channel,
and annihilation cross section $\langle\sigma v\rangle$), and an expected distribution of J-factors  determined from DM N-body simulations (described in section \ref{sec:sim-jfactor}), 
we estimate the expected distribution of DM annihilating subhalos that can be detected by {\Fermi} LAT for a certain observation time and instrument response functions (IRFs). 
The procedure is as follows:
\ben
\item
We start with a distribution $p(J) = dN / dJ$ of J-factors expected from N-body DM simulations (cf. figure~\ref{fig:Jfactors} in section~\ref{sec:sim-jfactor}). We sample the values of $J$ from $p(J)$. 
\item
Given a J-factor and DM parameters
$\btheta_{\rm DM} = \{\langle\sigma v\rangle, {\mdm}, {\rm ch.}\}$, 
we determine the expected gamma-ray spectrum as
\begin{equation}
\label{eq:DM_spec}
\phi_{\rm DM}(E;\btheta_{\rm DM}, J) = \frac{J}{4\pi} \frac{\langle \sigma v \rangle}{2\mdm^2} \left.\frac{dN_\gamma}{dE}\right|_{\rm ch.},
\end{equation}
where $dN_\gamma/dE$ is the theoretical spectrum of gamma rays produced in an annihilation of a pair of DM particles from ref.~\cite{2024JCAP...03..035A}. 
\item
We place DM subhalos at random positions in the sky with J-factors sampled in step 1 and with the spectra defined in eq.~(\ref{eq:DM_spec}).
\item 
For the diffuse gamma-ray background, we use the official \Fermi-LAT diffuse model template employed in the construction of the 4FGL-DR4 source catalog (\verb|gll_iem_v7|) and the corresponding isotropic background.
\item
We fit the gamma-ray data around each of the simulated DM subhalos with $|b| > 10^\circ$,
where we use the log-parabola model in eq.~(\ref{eq:log_par}) with fitting parameters $\{\log_{10}\phi, \alpha,\beta\}$. 
The fit is performed within a circular region centered on the target DM source. The radius of this region is energy-dependent, corresponding to the 95\% containment angle of the PSF, and is constrained to the range $[0.5^\circ, 5^\circ]$. This range is chosen to ensure sufficient statistics in the highest energy bins while simultaneously avoiding excessive background contamination in the lowest energy bins.
We use the same Gaussian priors for the $\beta$ parameter as in the 4FGL-DR4 catalog with mean 0.1 and sigma 0.3~\cite{2023arXiv230712546B}.
For details see appendix~\ref{app:sim-ps}.
\item
We keep DM subhalos detected with significance TS~$>25$, where ${\rm TS}=-2\ln {\cal L}/{\cal L}_*$ is the likelihood-ratio test statistic between the models with and without the subhalo. Such a threshold corresponds to the significance for PS detection in the {\Fermi}-LAT catalogs.
The fraction of subhalos passing the significance cut defines the detection efficiency $\varepsilon_{\rm eff}(\phi_{\rm DM})$, which is a function of the DM flux. 
\item
In order to get a good estimate of the DM distribution, $p_{\rm DM}(\log_{10} \phi_{\rm DM}, \alpha,\beta|\btheta_{\rm DM})$, we simulate realizations of the gamma-ray sky until we detect at least 20,000 DM subhalos across all simulations or the number of simulations reaches 1,000. The latter can happen for parameters, where DM subhalos have very small detection probability.%
\footnote{The limit of a thousand simulations is needed to avoid infinite iterations for low-efficiency parameter combinations.
For details on simulation and detection of DM subhalos, see appendix~\ref{app:sim-ps}.
}
\een
In order to reduce the computation time, we perform simulations in flux bins.
At first, we determine the distribution of $\alpha$ and $\beta$ parameters, given the flux $\phi_{\rm DM}$ and DM parameters $\btheta_{\rm DM}$, $p_{\rm DM}(\alpha,\beta|\phi_{\rm DM};\btheta_{\rm DM})$, by modeling the distribution of recovered $\alpha$ and $\beta$ parameters in this flux bin (see appendix~\ref{app:sim-ps} for details).
Then we multiply this distribution by the distribution of fluxes, which depends on the distribution of J-factors
$n(\phi_{\rm DM}) = \frac{dN}{d\phi_{\rm DM}} (\phi_{\rm DM};\btheta_{\rm DM}) = \frac{dN}{dJ} \frac{dJ}{d\phi_{\rm DM}}$
and by detection efficiency as a function of flux $\varepsilon_{\rm eff}(\phi_{\rm DM})$ to determine the probability density times the DM subhalo weight in eq.~(\ref{eq:mixture}):
\begin{equation}
\pi_{\rm DM}(\btheta_{\rm DM}) p_{\rm DM}(\bx;\btheta_{\rm DM}) = 
\frac{1}{N^{\rm ROI}_{\rm unas}}
p_{\rm DM}(\alpha,\beta|\phi_{\rm DM};\btheta_{\rm DM}) 
n(\phi_{\rm DM})
\varepsilon_{\rm eff}(\phi_{\rm DM}),
\label{eq:pDM}
\end{equation}
where $N^{\rm ROI}_{\rm unas}$ is the number of unassociated sources in the ROI.

We note that given the DM parameters, the contribution of DM subhalos to unassociated sources is fixed.
In particular, the number of subhalos in our ROI is
\be 
\label{eq:N_ROI}
N^{\rm ROI}_{\rm sub}(\btheta_{\rm DM}) =
\int_{\rm ROI} d\alpha ~ d\beta ~ d\phi_{\rm DM}~p_{\rm DM}(\alpha,\beta|\phi_{\rm DM};\btheta_{\rm DM}) ~n(\phi_{\rm DM}) \varepsilon_{\rm eff}(\phi_{\rm DM}),
\ee
while the DM weight $\pi_{\rm DM}$ in eq.~(\ref{eq:mixture}) is 
\be
\pi_{\rm DM}(\btheta_{\rm DM}) = \frac{N^{\rm ROI}_{\rm sub}(\btheta_{\rm DM})}{N^{\rm ROI}_{\rm unas}}.
\ee

\begin{figure}
    \centering
    \includegraphics[width=0.45\linewidth]{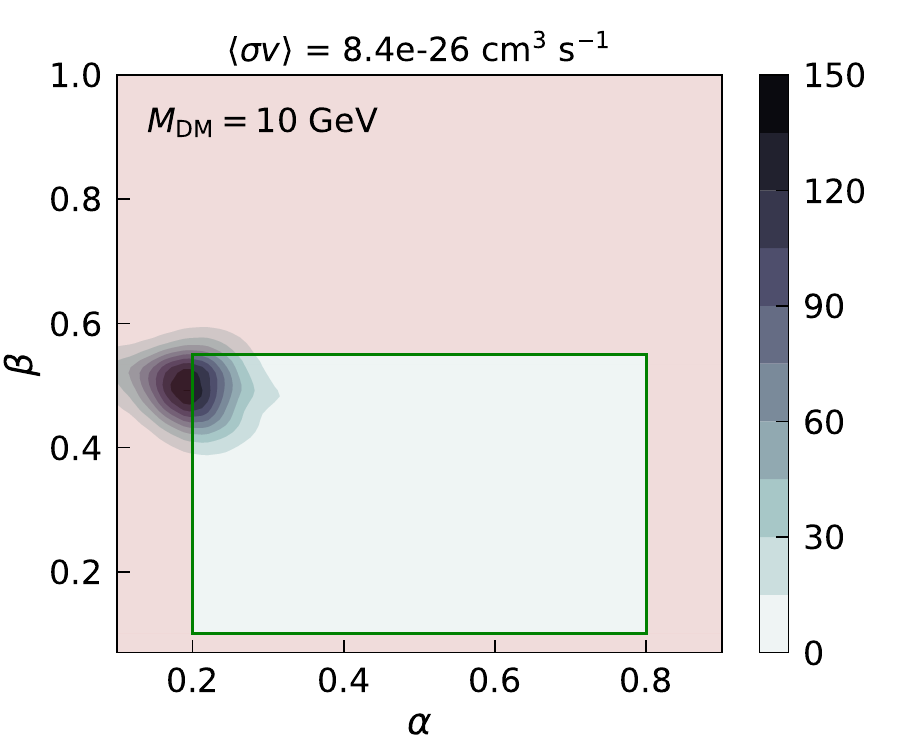}
    \includegraphics[width=0.45\linewidth]{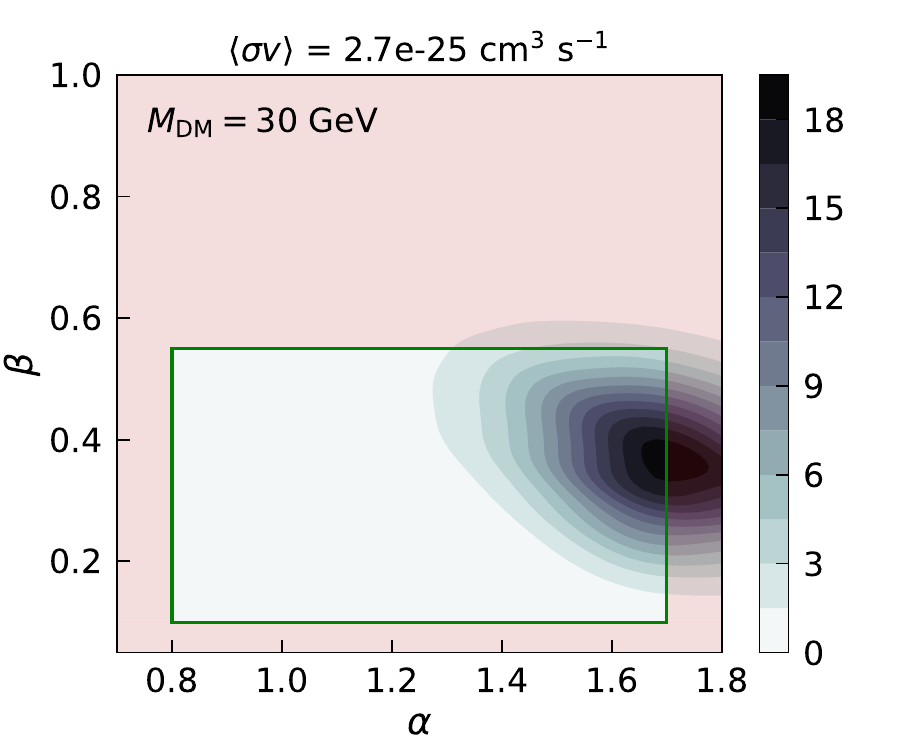}
    \includegraphics[width=0.45\linewidth]{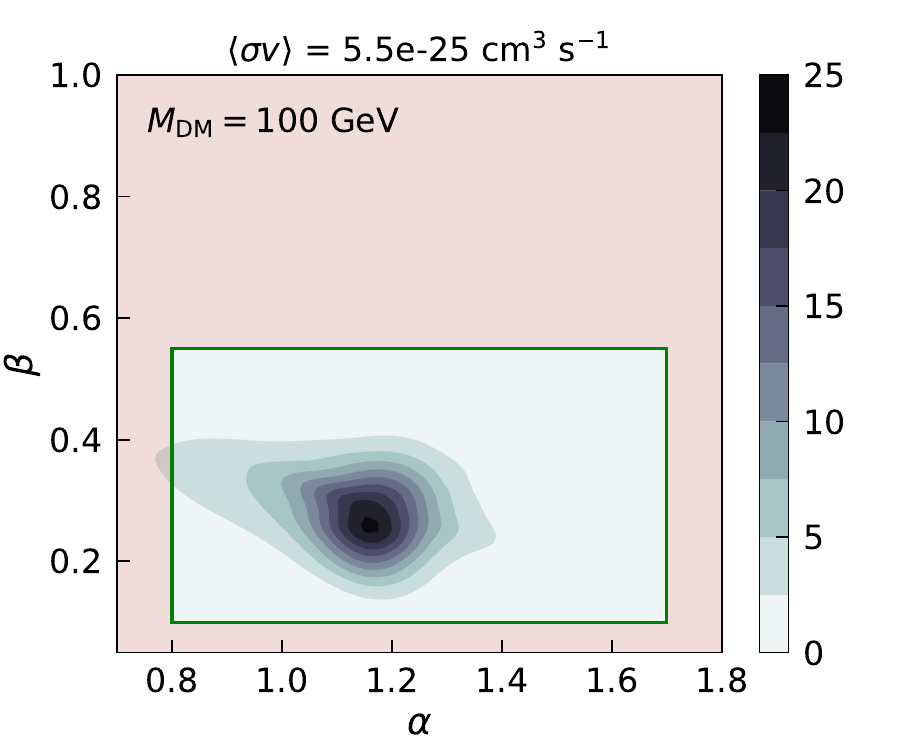}
    \includegraphics[width=0.45\linewidth]{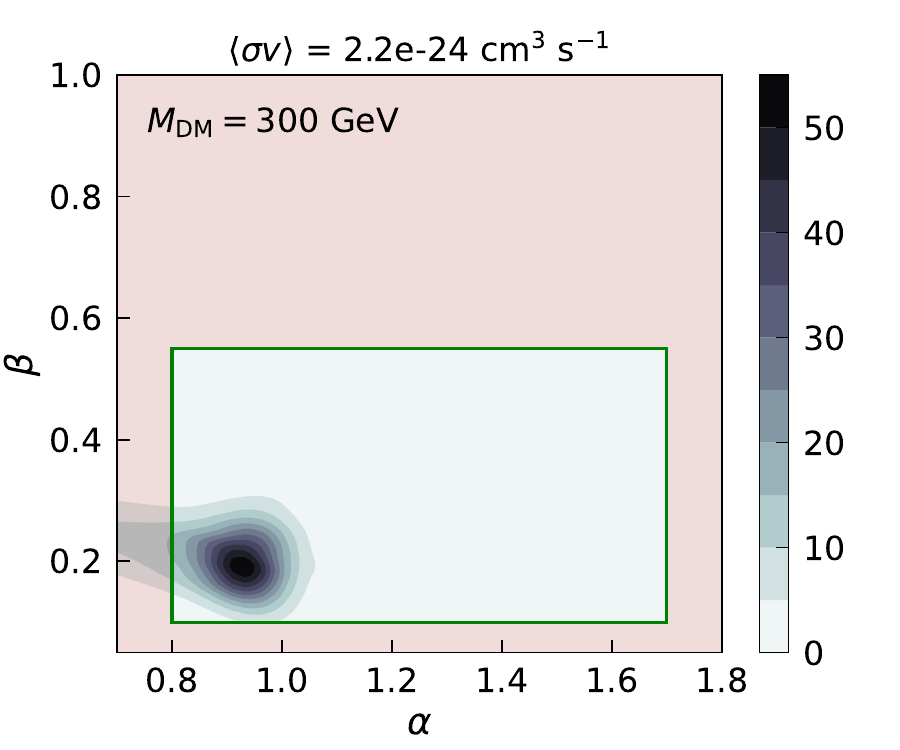}
    \includegraphics[width=0.45\linewidth]{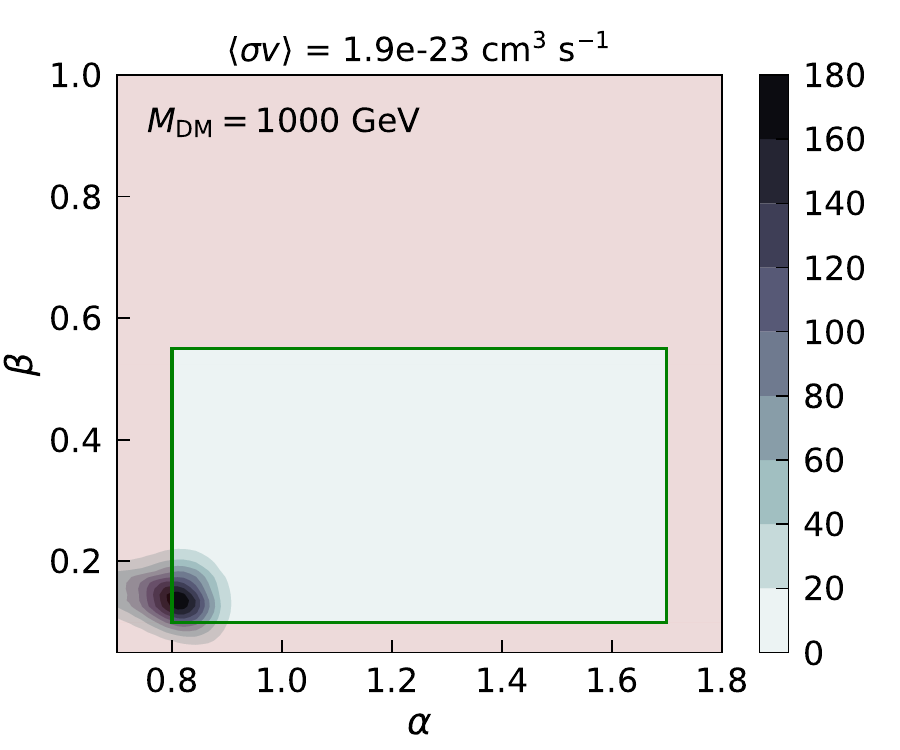}
    \caption{DM distribution at different DM masses, $b\overline{b}$ annihilation channel, integrated in flux for visualization purposes. The $\langle\sigma v\rangle$ values are chosen close to our bound values. 
    The green boxes indicate the parameter ranges (ROI) used for this analysis and define the truncation region for the distribution. The color bars show the value of the DM subhalo PDF (eq. \ref{eq:pDM}), normalized to unit integral within the ROI.
    }
    \label{fig:dmdist-all}
\end{figure}

In figure~\ref{fig:dmdist-all} we show the distributions of DM subhalos in $\alpha$ and $\beta$ parameters (we integrate the distributions in the flux dimension for presentation purposes) for different DM masses at some characteristic $\langle\sigma v\rangle$ values. 
We note that the DM distributions are not fully contained inside the  ROI that we selected in this analysis, which is indicated as a green box in the plots. 

In particular, a significant fraction of the DM subhalo distribution is truncated for DM masses of 10 GeV, 30 GeV, and 1 TeV.
{Although choosing a larger ROI that covers a larger fraction of the DM subhalo distribution could increase statistics, our assumption that the covariate shift can be modeled by a monotonous function would not be valid anymore. Introducing a more flexible covariate shift model to account for this would require additional parameters, potentially leading to larger uncertainty in $\pi_{\rm DM}$ due to correlations with these new parameters (i.e., a larger parameter covariance matrix) and, consequently, weaker limits. Moreover, the gain from larger statistics would be limited by the significant background of astrophysical sources outside of our current ROI, particularly unassociated millisecond pulsars with $\alpha > 0.7$ and $\beta \sim 0.4$ (upper right panel of figure~\ref{fig:dmdist-all}). Conversely, regions with small $\alpha$ values (bottom panel) contain very few astrophysical sources, necessitating extrapolation that carries large systematic uncertainties (see the wide uncertainty band for small alpha values in the upper left panel in figure~\ref{fig:dist_ratio}). Finally, our limits on DM annihilation are conservative: positive residuals after fitting the data with astrophysical components can be absorbed by the DM component, resulting in a positive best-fit $\pi_{\rm DM}$ and weaker upper limits than in the $\pi_{\rm DM} = 0$ case.}

\begin{figure}
    \centering
    \resizebox{0.95\textwidth}{!}{
    \includegraphics[width=0.49\linewidth]{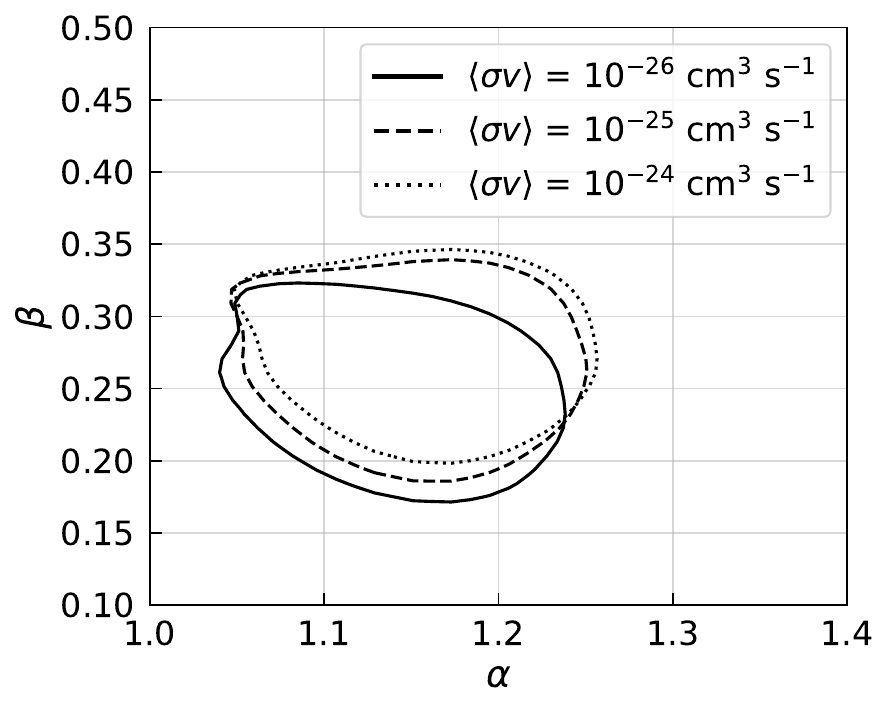}
    \includegraphics[width=0.49\linewidth]{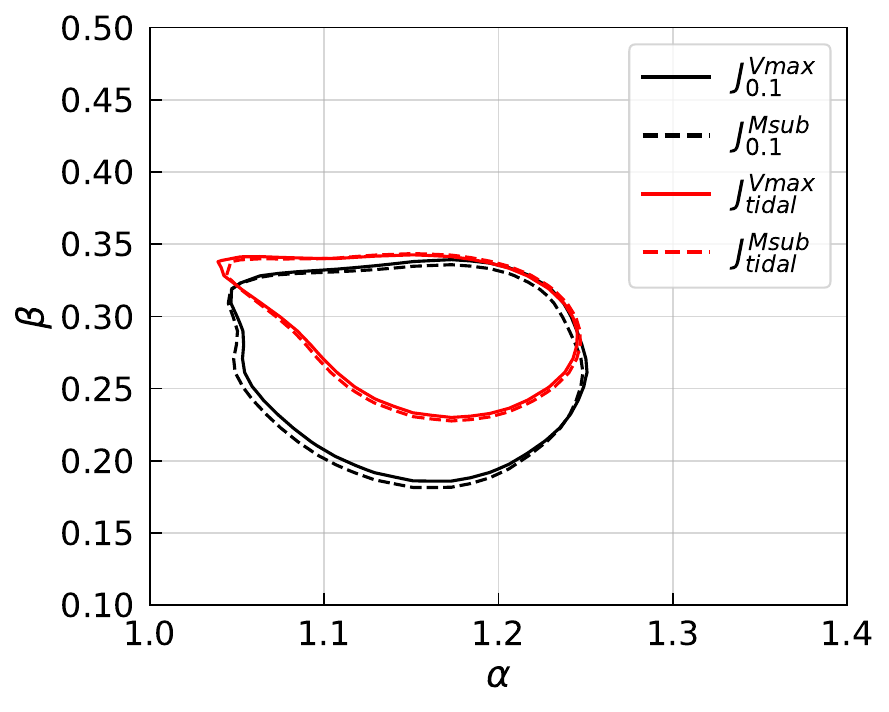}
    }
    \caption{Contour plots at half maximum of the DM subhalo distribution for different values of $\sigmav$ and J-factor models at $\mdm$ = 100 GeV. \panel{Left panel}: contour lines for different choices of $\sigmav$ for $J_{0.1}^{Vmax}$. 
    \panel{Right panel}: contour lines for different J-factor models for $\sigmav = 10^{-25}$ ${\rm cm}^{3} \, {\rm s}^{-1}$.}
    \label{fig:DM_dist_countour}
\end{figure}

In figure~\ref{fig:DM_dist_countour} we show the contour levels for the half maximum of the DM subhalo distribution for different choices of $\sigmav$ and J-factor models for $\mdm$ = 100 GeV. 
The distribution shifts to higher values of $\beta$ for larger $\sigmav$ values. Analogously, choosing $J_{tidal}$ shifts the distribution to higher $\beta$ values compared to $J_{0.1}$.
This is to be expected, since for faint sources (small $\sigmav$ or small J-factors) the fit has large statistical uncertainties and the best-fit value of the $\beta$ parameter is more affected by the priors on $\beta$ than in case of bright sources.

\section{Mixture model of gamma-ray sources and limits on DM annihilation}
\label{sec:results}

\subsection{Model optimization}
\label{sec:model-opt}

Our general aim is to either detect a population of DM subhalos among the \Fermi-LAT unassociated sources or to put an upper limit on $\langle\sigma v\rangle$ for different DM masses. For a fixed DM mass and annihilation channel, we determine the best-fit cross section $\langle\sigma v\rangle^*$ (optimized simultaneously with the other parameters) by maximizing the full likelihood built from the model in eq.~(\ref{eq:mixture}), defined as:
\be
\label{eq:L}
{\cal L}(\langle\sigma v\rangle, \pi_k, \btheta_{\rm cov}) = \prod_{i=1}^{N^{\rm ROI}_{\rm unas}} \tilde{p}_{\rm unas}(\bx_i|\sigmav,\pi_k, \btheta_{\rm cov};\mdm,{\rm ch.}).
\ee

We maximize eq.~(\ref{eq:L}) with the Expectation-Maximization (EM) algorithm, which is efficient for the cases of probabilistic mixture models~\cite{10.1162/089976602753284446_Saerens}, see appendix~\ref{app:EM} for more details.
From the ML point of view our approach is a semi-supervised learning algorithm.

Indeed, in supervised learning the model is fixed by the training data (including the PDFs for the different classes), while in unsupervised learning, the model is determined solely based on target data.
In our case, we use the associated sources (training data) to determine PDFs of the two classes of associated sources (which is similar to supervised learning), but we further modify the PDFs to account for the possible dataset shift between training and target data and add a new class of sources (which is similar to unsupervised learning).

We optimize the likelihood for a set of DM mass values: $\mdm$ = 10, 30, 100, 300, 1000 GeV, and determine, for each $\mdm$, the upper bound on $\sigmav$ at 95\% confidence level (CL) by profiling the test-statistic 
\be
{\rm TS}(\sigmav) = -2\ln \frac{{\cal L}(\sigmav)}
{{\cal L}_{\rm max}}, 
\label{eq:TS}
\ee
where ${\cal L}(\sigmav)$ is the likelihood optimized over the rest of parameters $\pi_k$ and $\btheta_{\rm cov}$, and ${\cal L}_{\rm max}={\cal L}(\sigmav^*,\pi_k^*,\btheta_{\rm cov}^*)$ is the value of the likelihood for the best-fit parameters. 
Following the Wilks' theorem, we assume that TS follows the $\chi^2$ distribution. 
We compute the upper bound on $\sigmav$ as the value corresponding to the 95\% coverage probability (one-sided), which for one degree of freedom corresponds to
TS = 3.84.

\begin{figure}[t]
    \centering
    \resizebox{0.99\textwidth}{!}{
    \hspace{-3mm}
    \includegraphics[width=0.33\linewidth]{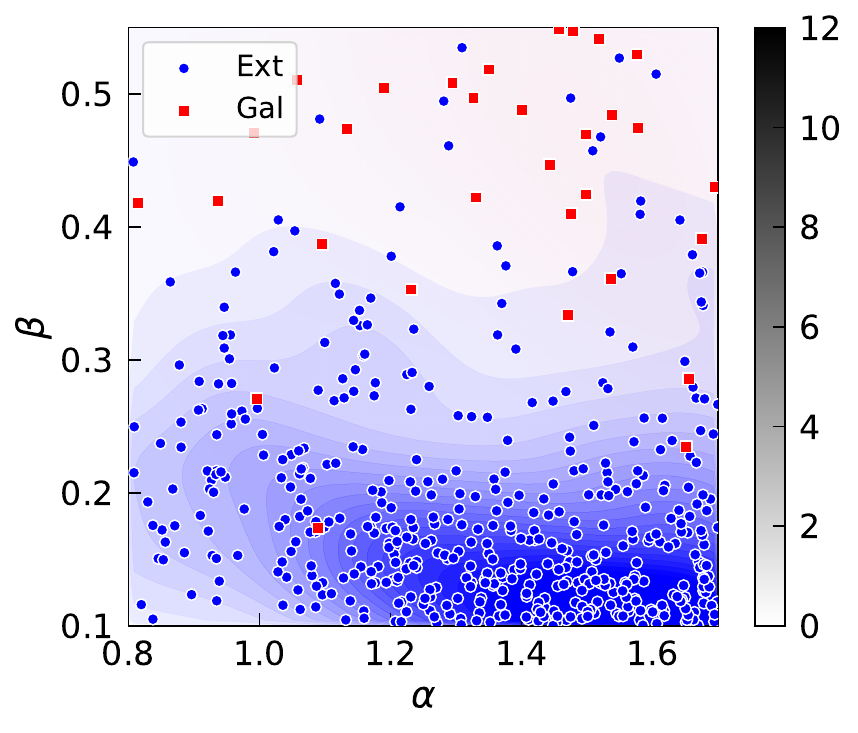}
    \hspace{-3mm}
    \includegraphics[width=0.33\linewidth]{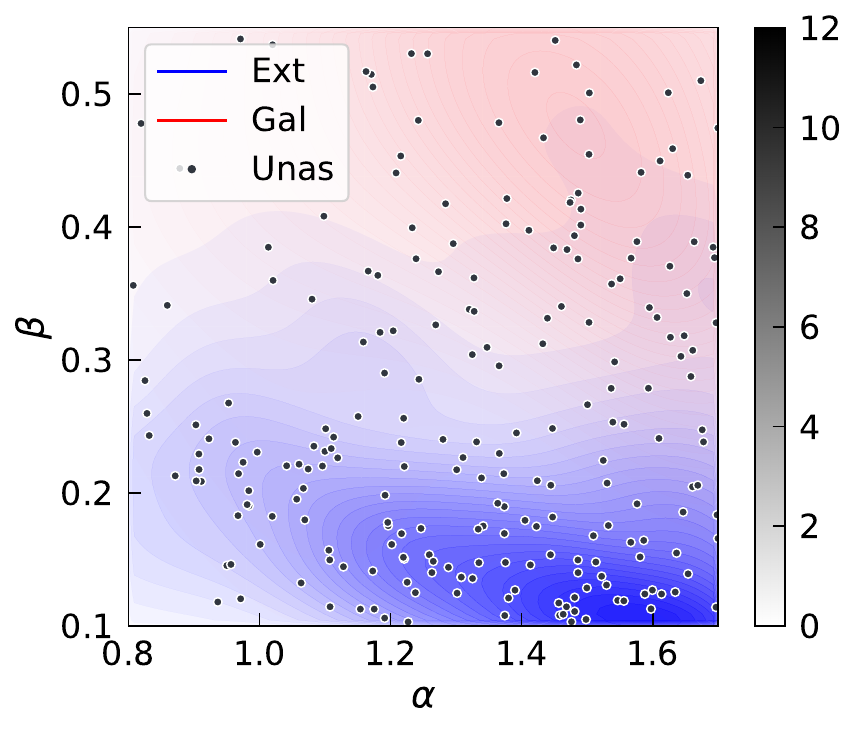}
    \hspace{-3mm}
    \includegraphics[width=0.33\linewidth]{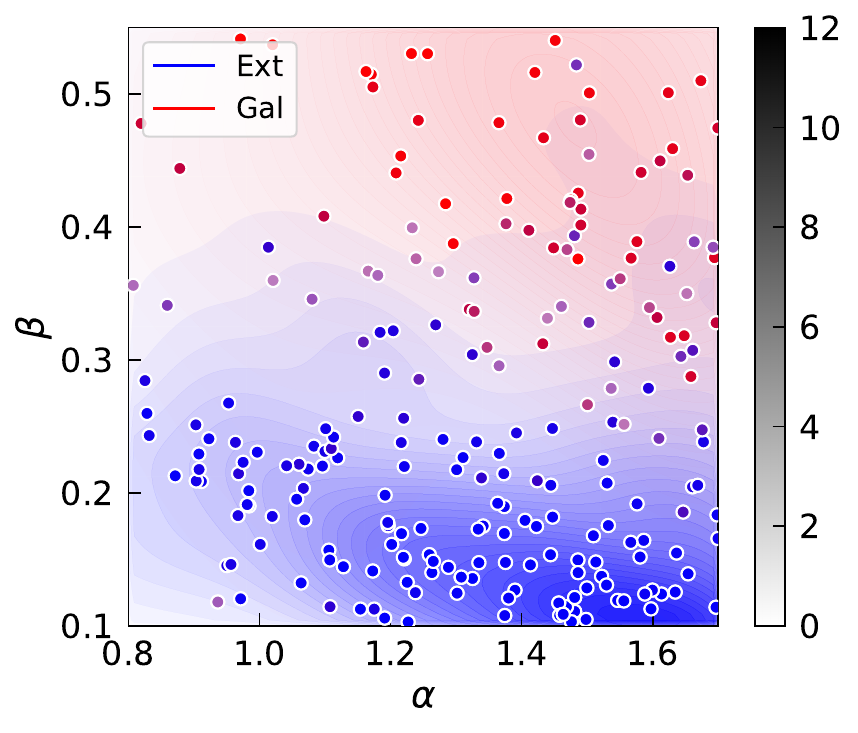}
    }
    \caption{
    Distributions of sources in the plane of the spectral parameters ($\alpha$, $\beta$) for $E_0$ = 1~GeV.
    \panel{Left panel}: PDFs of Galactic and extragalactic sources weighted by the corresponding class fractions. The sources are colored according to the catalog class.
    \panel{Middle panel}: best-fit mixture model for the unassociated sources. The background colors show the individual components of the astrophysical-only model from eq.~(\ref{eq:mixture}), where the DM subhalo contribution is set to zero. Specifically, we plot the decomposition of the mixture model into class-conditional PDFs weighted by their best-fit class prevalence $\pi_k p(x | k)$ for the extragalactic (blue) and Galactic (red) populations. For visualization, the distributions are integrated over the flux dimension. The color bar represents the value of the resulting probability density (shown in gray to represent both red and blue colors).
    \panel{Right panel}: Class probabilities for unassociated 4FGL-DR4 sources (for $\rm E_0 = 1 \,GeV$), on top of the mixture distribution (same as the middle panel). The color scale is proportional to the class probabilities, according to the formula ${\rm color}(x_i)={\rm red}\cdot p(C_k= {\rm gal}|x_i) + {\rm blue}\cdot p(C_k={\rm ext}|x_i)$, e.g., a purple-colored source is a source for which the posterior class probabilities are similar and the classification is ambiguous. 
    }
    \label{fig:mixtures}
\end{figure}

We maximize the likelihood of the model in eq.~(\ref{eq:L}) to determine the best-fit parameters.
At first, we fix the DM contribution to be zero and determine the best-fit model with the known astrophysical sources only.
For reference, we show the model for the associated sources (no fitting) in the left panel of figure~\ref{fig:mixtures} and the best-fit model for unassociated sources in the middle and right panels.
In these plots, we display the distributions evaluated in the analysis ROI for $\alpha$ and $\beta$ parameters integrating over the flux coordinate for visualization purposes. We decompose the mixture model of eq.~(\ref{eq:mixture}) and associate the blue color with the extragalactic component and the red color with the Galactic component, both respectively weighted by their prior weight $\pi_k$ to highlight their relative contribution. The intensity of the color bar corresponds to the value of the component likelihood weighted by the prior weight, $\pi_k p(x | k)$, and a darker, more intense color is associated with regions where there is a higher probability of observing a certain source class. We will follow this convention for all upcoming distribution plots.

\begin{figure}[t]
    \centering
    \resizebox{0.95\textwidth}{!}{
    \includegraphics[width=0.49\linewidth]{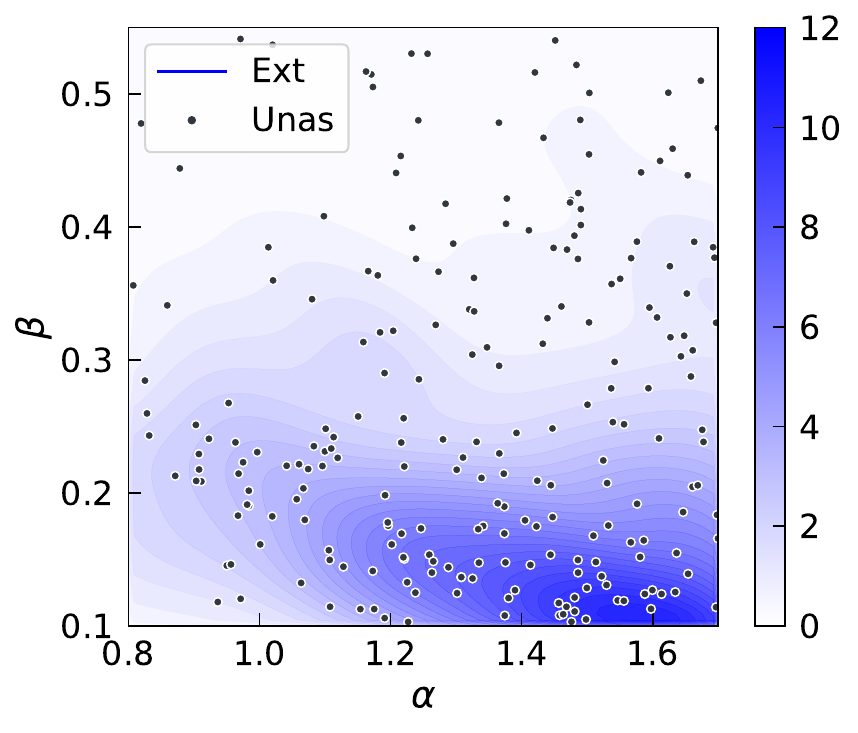}
    \includegraphics[width=0.49\linewidth]{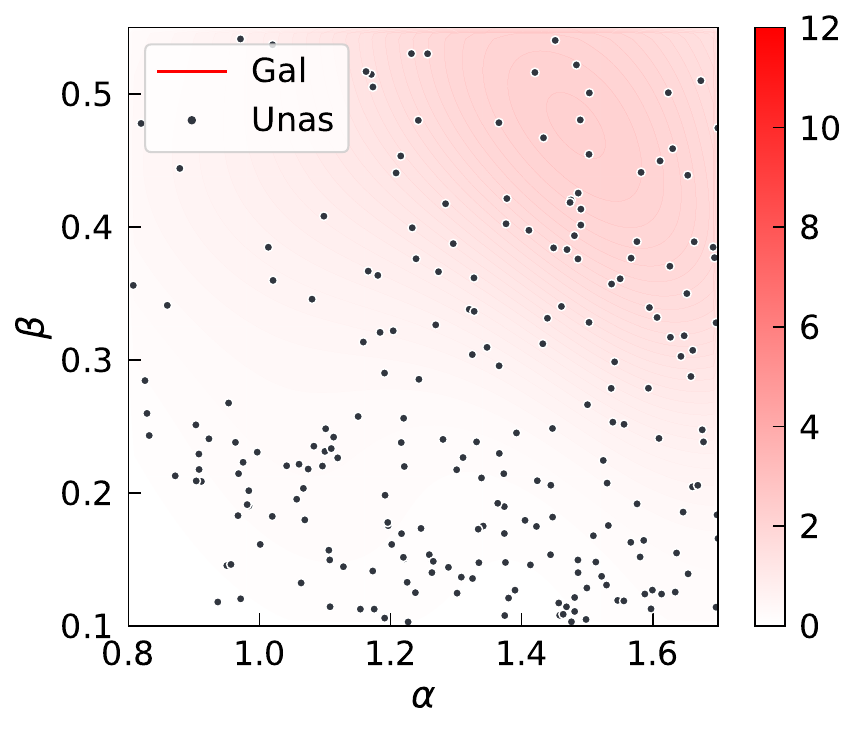}
    }
    \caption{Decomposition of the mixture model for unassociated sources with two classes (Galactic and extragalactic) for the ROI cut associated with $E_0 = 1$ GeV. The model is the same as in the middle and right panels of figure~\ref{fig:mixtures}.  
    \panel{Left panel}: extragalactic component. \panel{Right panel}: Galactic component.
    }
    \label{fig:unas_mixture_astro}
\end{figure}

We note that the extragalactic sources are dominated by AGNs, which usually have close to a power-law spectra (small $\beta$ values).
The Galactic sources often have curved spectra (large $\beta$ values).
The relative contribution of Galactic sources is larger for the unassociated sources compared to the associated ones (brighter red color in the right panel compared to the left panel in figure~\ref{fig:mixtures}).
The model on the middle and right panels of figure~\ref{fig:mixtures} is obtained by optimizing the prior and covariate shift parameters, $\pi_k$ and $\btheta_{\rm cov}$.
We display the optimized Galactic and extragalactic components separately in figure~\ref{fig:unas_mixture_astro}.
The corresponding best-fit parameters $\pi_k$ (the class prevalence) are presented in table~\ref{tab:class_prevalence_astro}. 
As qualitatively discussed above, the expected contribution of Galactic sources to unassociated sources at high latitudes (29\%) is significantly larger than the fraction of associated Galactic sources (6\%) in the ROI.
The right panel of figure~\ref{fig:mixtures} also shows the model's predictions for class probabilities of unassociated sources. 
{
We compare the relative importance of prior and covariate shift parts of the model in appendix~\ref{app:prior_vs_cov}, where
we find that the prior shift plays a more important role than the covariate shift.
In particular, for the model without the DM component, we find that the prior shift is preferred at $7\sigma$ significance (for one additional degree of freedom) compared to a model without either prior or covariate shifts, while the model with the covariate shift is preferred at $4.5\sigma$ significance (with 6 additional degrees of freedom). The model with both prior and covariate shifts is preferred at $6.7\sigma$ significance (with 7 degrees of freedom), i.e., most of the improvement is, indeed, due to the prior shift.
}

\begin{table}[t]
\centering
\begin{tabular}{|rccccc|}
\toprule
type & $\rm N_{ext}$ & $\rm N_{Gal}$ & \% Ext & \% Gal & $\rm N_{tot}$\\
\midrule
assoc & 466 & 31 & 94\% & 6\% & 497\\
unas & 160.6 & 66.4 & 71\% & 29\% & 227\\
\bottomrule
\end{tabular}
\caption{Class prevalence for Galactic and extragalactic sources in the ROI. 
For the associated sources (second row) the numbers are derived from the 4FGL-DR4 catalog. 
For the unassociated sources (third row), the numbers correspond to the best-fit model with astrophysical components only (using $E_0 = 1$ GeV). 
The percent of galactic (\% Gal) and extragalactic (\% Ext) sources are given by the best-fit parameters $\pi_k$ in eq.~(\ref{eq:mixture}) for the two astrophysical components.
The expected number of Galactic and extragalactic sources among the unassociated ones is derived as $\pi_k \times N_{\rm tot}$, where $N_{\rm tot} = 227$ is the total number of unassociated sources in the ROI.
}
\label{tab:class_prevalence_astro}
\end{table}

\subsection{Dark matter limits}
\label{sec:dm-limits}

\begin{figure}
    \centering
    \includegraphics[width=0.49\linewidth]{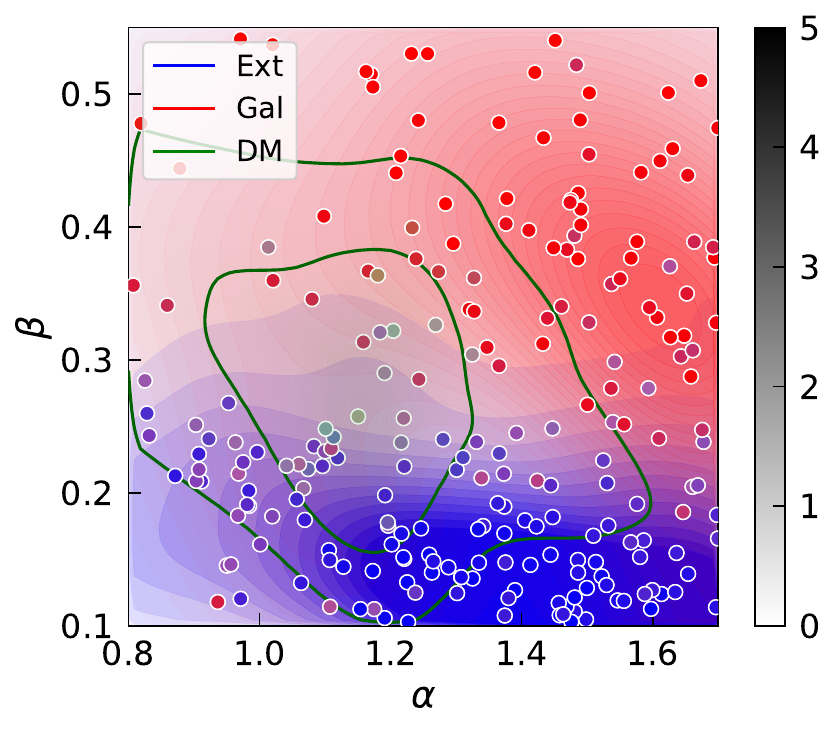}
    \caption{Mixture model for unassociated sources including the DM component with ${\mdm} = 100$ GeV, following the same conventions as in figure~\ref{fig:mixtures}.
    The green component represents the best-fit DM distribution for $\sigmav=3.4 \cdot 10^{-25} {\rm cm}^3{\rm s}^{-1}$. The green lines represent the 68\% and 95\% containment levels for the DM distribution.
    The background colors represent the class-conditional distributions weighted by their prior weight: $\pi_kp(x|k)$.
    Sources are colored according to their posterior class probability following the formula ${\rm color}(x_i)={\rm red}\cdot p(C_k= {\rm gal}|x_i) + {\rm blue}\cdot p(C_k={\rm ext}|x_i) + {\rm green}\cdot p(C_k={\rm DM}|x_i)$.
    }

    \label{fig:mixture_DM_tot}
\end{figure}

\begin{figure}
    \centering
    \resizebox{0.99\textwidth}{!}{
    \hspace{-3mm}
    \includegraphics[width=0.32\linewidth]{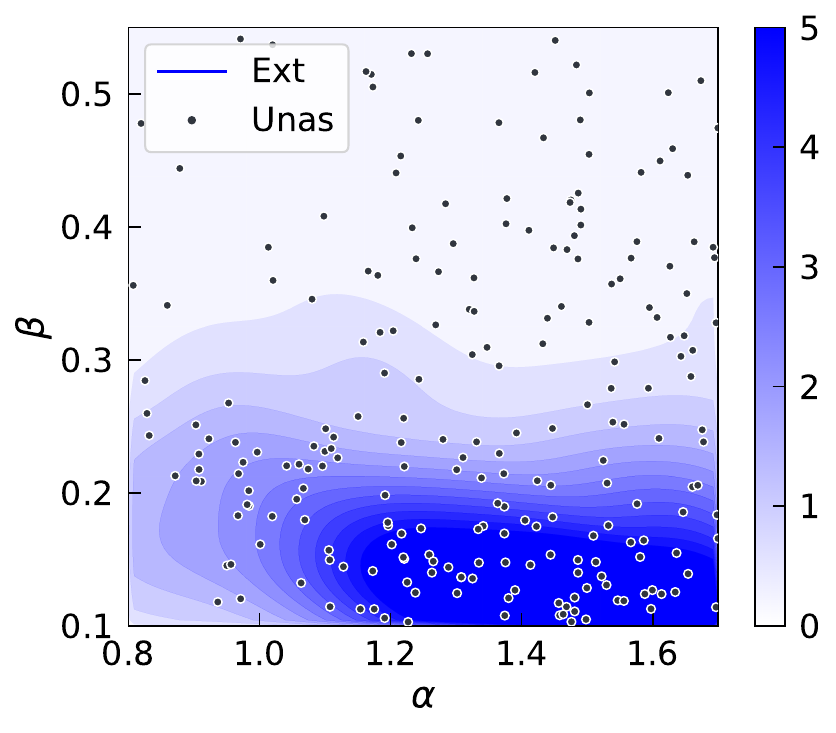}
    \hspace{-3mm}
    \includegraphics[width=0.32\linewidth]{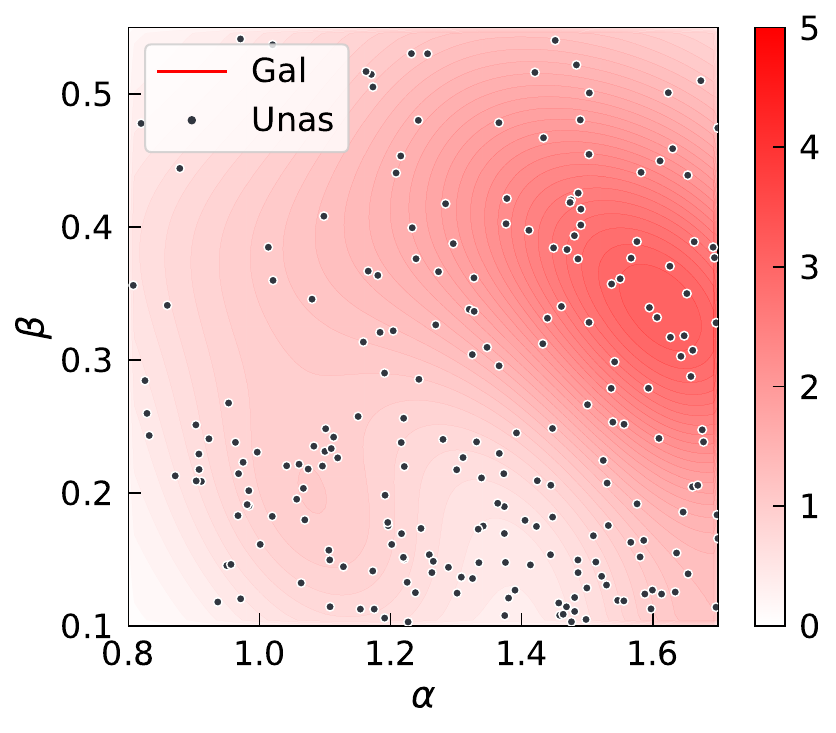}
    \hspace{-3mm}
    \includegraphics[width=0.32\linewidth]{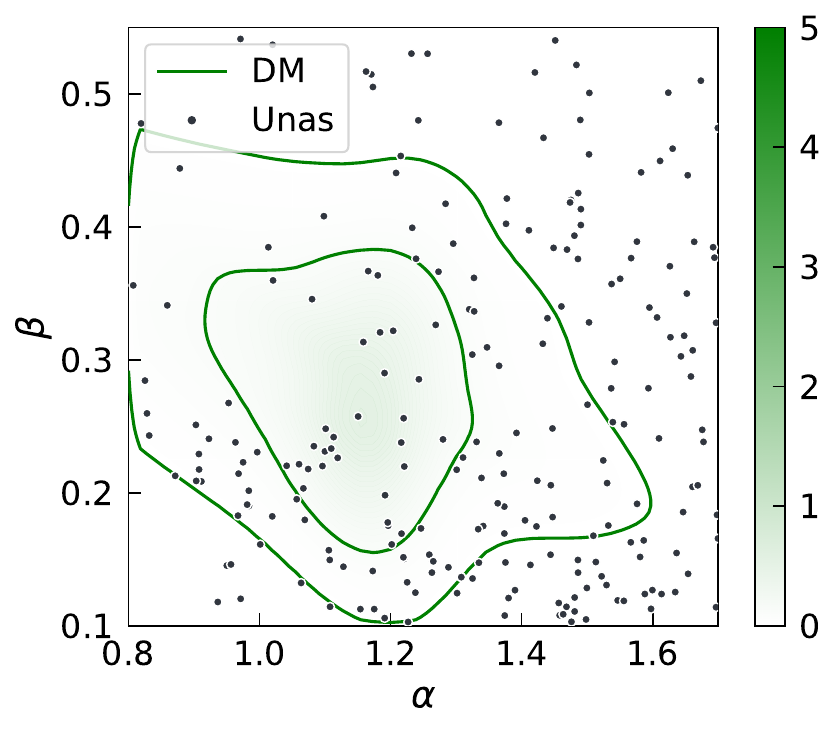}
    }
    \caption{Components of the mixture model in figure~\ref{fig:mixture_DM_tot}, following the same conventions as in figure~\ref{fig:mixtures}.
    }
    \label{fig:mixture_DM_components}
\end{figure}

The best-fit model including the DM component with DM mass of 100 GeV and $\sigmav=3.4 \cdot 10^{-25} {\rm cm}^3{\rm s}^{-1}$ is shown in figure~\ref{fig:mixture_DM_tot}.
We simultaneously fit the parameters $\pi_k$, $\btheta_{\rm cov}$ and $\sigmav$.
The corresponding components are presented in figure~\ref{fig:mixture_DM_components}.
The DM component is shown by the green color.%
\footnote{The class probabilities for unassociated sources in the astrophysical only model, as well as DM subhalo class probabilities for the different DM masses, are available online at \href{https://doi.org/10.5281/zenodo.17463823}{https://doi.org/10.5281/zenodo.17463823}.
}

Provided that all DM subhalos have the same gamma-ray spectra, the corresponding distribution of DM subhalos is much better localized in the $\alpha$ and $\beta$ plane than the extragalactic and Galactic components.
The spread in the DM subhalo distribution comes from statistical uncertainties in the reconstruction of spectral parameters.
We note that not only the overall normalization but also the shape of the distribution of DM subhalos in the right panel of figure~\ref{fig:mixture_DM_components}
changes for different annihilation cross sections and J-factor models (cf. figure~\ref{fig:DM_dist_countour}).
The reason is that the distribution depends on the relative contribution of subhalos with different J-factors, e.g., smaller J-factors correspond to proportionally smaller flux and larger uncertainties, which affect the shape of the distribution.
Nevertheless, for larger $\sigmav$ values the expected contribution of DM subhalos also increases.

We present the details of the best-fit model for the DM subhalo component (including upper limits in the annihilation cross section)
in table~\ref{tab:Nsub_values}.
Columns 2 to 6 show the best-fit model, e.g., the best-fit DM subhalo prevalence, the expected number of subhalos within our region of interest ($N_{\rm subhalo}^{\rm ROI}$) and for the full range of parameters ($N_{\rm sub}^{\rm total}$), as well as the best-fit thermally-averaged annihilation cross-section ($\sigmav$). The value of $\Delta {\rm TS}_0 = -2(\log\mathcal{L}_0 - \log\mathcal{L}_{\rm max})$ quantifies the preference for the best-fit model over a baseline astrophysical-only scenario. 
For all tested models, the significance is less than 2 sigma.
We notice that the largest possible contribution from DM subhalos is for $\mdm = 30$ GeV (although still not significant). 

The expected number of DM subhalos of 36.2 for this DM mass in the full range of parameters is comparable to the 16 DM subhalo candidates reported in ref.~\cite{2019JCAP...07..020C},
once the improved sensitivity of the 4FGL-DR4 catalog (14 years of observations) relative to the 3FGL catalog (8 years of observations) is taken into account.
To verify this, we first consider the case where the DM subhalo-like sources are astrophysical in origin. In the Euclidean regime of constant volume density, the source counts scale as $N(> S) \propto S^{-1.5}$ as a function of flux, and the detection threshold scales as $S \propto t^{-0.5}$. Under these assumptions, 36.2 subhalo-like sources in 14 years of observations (4FGL-DR4, full range of parameters) correspond to approximately 23.8 sources in 8 years of observations (3FGL catalog, ref.~\cite{2019JCAP...07..020C}).
Since 16 is within about $1.6\sigma$ of 23.8, the two methods yield consistent estimates of the number of possible DM subhalos.
Alternatively, if the sources are DM annihilating subhalos, our baseline model has $dN/dJ \propto J^{-3}$ (black dashed line in figure~\ref{fig:Jfactors}), which gives $N_{\rm sub} \propto J^{-2} \propto t$, where we have used $S \propto J$ for fixed $\sigmav$.
In this case, 36.2 detectable subhalos in 14 years correspond to $36.2 \times 8 / 14 \approx 20.7$ subhalos in 8 years, which is also consistent with the 16 subhalos assumed in ref.~\cite{2019JCAP...07..020C}.
We also notice that for $\mdm = 30$~GeV,  the expected contribution from Galactic sources (mostly millisecond pulsars) is the smallest.
This is to be expected, since the spectrum of DM annihilation in the $b\bar{b}$ channel is similar to the spectra of pulsars at this DM mass.
As a result, there may be a degeneracy between the DM subhalos and Galactic sources for this DM mass.
Provided that we do not find significant evidence for a signal from DM, we proceed to set conservative upper bounds on $\sigmav$ shown in the ``upper bound" section of the table.

{
In order to estimate the overall goodness of fit of the model in eq.~(\ref{eq:mixture}) without DM to the data,
we compare in appendix~\ref{app:model_goodness} the log likelihood of the fit to the data with the distribution of log likelihood values for the data generated from the best-fit model.
We find that the deviation of the log likelihood for the real data is less than $2\sigma$ smaller than the mean log likelihood value for the simulated data, i.e., the deviation is consistent with statistical fluctuations within $2\sigma$.
We also perform DM injection tests in appendix~\ref{app:DM_injection}.
We find that our method can recover the DM signal within statistical fluctuations: for all DM masses the  $\sigmav$ value for the injected signal is within the 68\% containment of the recovered $\sigmav$ values around the median recovered $\sigmav$.
}
{In addition, we determine the precision-recall curves for the classification of individual DM subhalos for 1000 simulations with injected DM signal (in case of $\mdm = 30$~GeV) and evaluate the average precision of our model in appendix \ref{app:pr_curve}.}

\begin{table}[t]
\centering
\resizebox{0.95\textwidth}{!}{%
\begin{tabular}{|l|ccccc|ccc|}
\toprule
 & \multicolumn{5}{c|}{\textbf{best-fit model}} & \multicolumn{3}{c|}{\textbf{upper bound}} \\ 
 $\mdm$ & \% DM & $N_{\rm subhalo}^{\rm ROI}$  & $N_{\rm subhalo}^{\rm total}$ & $\langle \sigma v \rangle$ & $\Delta {\rm TS}_0$ & $N_{\rm subhalo}^{\rm ROI}$ & $N_{\rm subhalo}^{\rm total}$ & $\langle \sigma v \rangle$ \\
  $\rm [GeV]$ & & & & [$\rm cm^3 \, s^{-1}$] & & & & [$\rm cm^3 \, s^{-1}$] \\
\midrule
10 & 0.0 & 0.00 & 0.00 & 0.0 & 0.0 & 1.91& 4.42& $5.6 \times 10^{-26}$\\
30 & 6.8 & 15.37 & 36.23 & $2.7 \times 10^{-25}$& 3.5 & 27.30 & 64.20 & $3.6 \times 10^{-25}$\\
100 & 2.4 & 5.45 & 6.21 & $3.4 \times 10^{-25}$ & 1.2 & 19.36 & 21.86 & $5.9 \times 10^{-25}$\\
300 & 2.0 & 4.59 & 7.66 & $1.1 \times 10^{-24}$ & 1.1 & 13.23 & 21.46 & $1.7 \times 10^{-24}$\\
1000 & 0.0 & 0.00 & 0.00 & 0.0 & 0.0 & 2.06&  8.61& $4.7 \times 10^{-24}$ \\
\bottomrule
\end{tabular}%
}
\caption{
Expected number of DM subhalos ($N_{\rm sub}$) for the \textbf{best-fit model} and for the \textbf{95\% confidence upper bound} on the annihilation cross-section ($\sigmav$). In both cases, we report the subhalo count within our ROI and the total count. The test statistic difference quantifies the significance of the best-fit model over an astrophysical-only model, $\Delta {\rm TS}_0 = -2(\log\mathcal{L}_0 - \log\mathcal{L}_{\rm max})$. The number of subhalos, $N_{\rm sub}$, is constrained to be non-negative.
}
\label{tab:Nsub_values}
\end{table}

\begin{table}[t]
\centering
\resizebox{0.95\textwidth}{!}{%
\begin{tabular}{|l|ccccc|ccc|}
\toprule
 & \multicolumn{5}{c|}{\textbf{best-fit model}} & \multicolumn{3}{c|}{\textbf{upper bound}} \\ 
 $\mdm$ & \% DM & $N_{\rm subhalo}^{\rm ROI}$  & $N_{\rm subhalo}^{\rm total}$ & $\langle \sigma v \rangle$ & $\Delta {\rm TS}_0$ & $N_{\rm subhalo}^{\rm ROI}$ & $N_{\rm subhalo}^{\rm total}$ & $\langle \sigma v \rangle$ \\
  $\rm [GeV]$ & & & & [$\rm cm^3 \, s^{-1}$] & & & & [$\rm cm^3 \, s^{-1}$] \\
\midrule
10 & 0.0 & 0.00 & 0.00 & 0.00 & 0.0 & 1.65 & 4.97 & $1.86 \times 10^{-26}$\\
30 & 0.0 & 0.00 & 0.00 & 0.00 & 0.0 & 12.41 & 26.80 & $1.61 \times 10^{-25}$\\
100 & 0.0 & 0.00 & 0.00 & 0.00 & 0.0 & 11.86 & 12.99 & $2.33 \times 10^{-25}$\\
300 & 4.3 & 10.87 & 16.67 & $8.79 \times 10^{-25}$ & 3.6 & 20.71 & 31.80 & $1.55 \times 10^{-24}$\\
1000 & 0.0 & 0.00 & 0.00 & 0.00 & 0.0 & 2.13 & 6.63 & $1.64 \times 10^{-24}$ \\
\bottomrule
\end{tabular}%
}
\caption{Same as Table \ref{tab:Nsub_values}, but for the $\rm J_{tidal}^{Msub}$ case.
}
\label{tab:Nsub_values_Msub}
\end{table}

\begin{figure}
    \centering
    \includegraphics[width=0.80\linewidth]{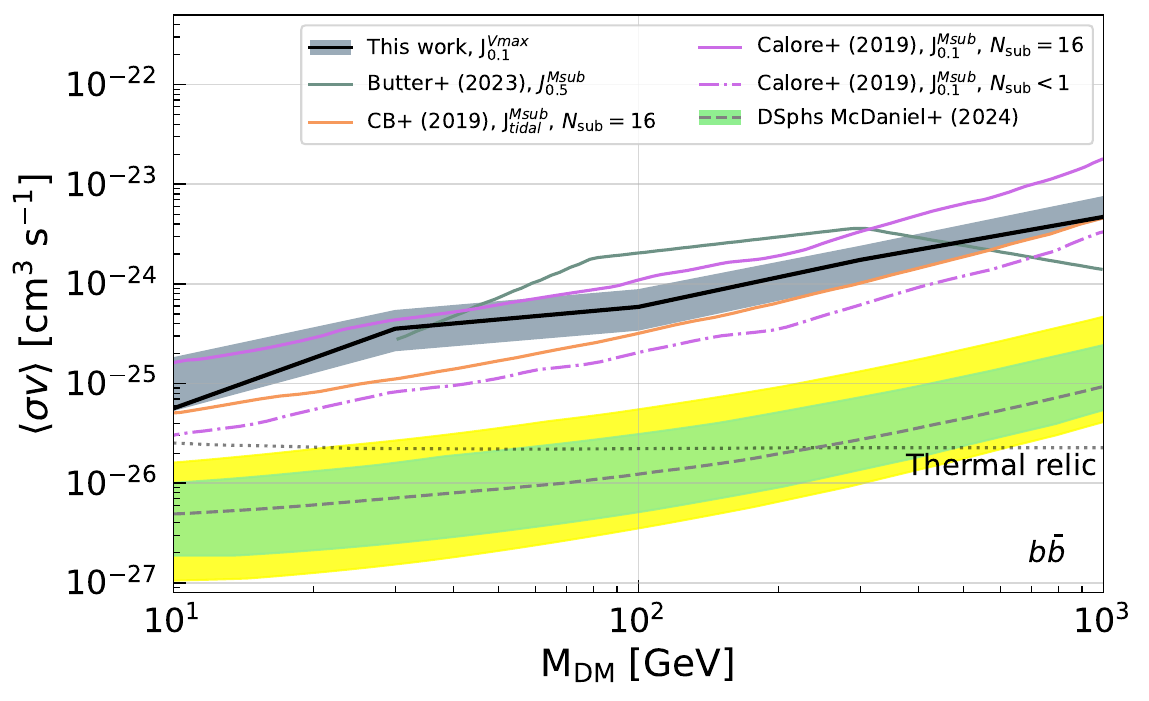}
    \caption{
    Upper bounds with 95\% confidence on the DM annihilation cross section $\sigmav$ in the $b\bar{b}$ channel. 
    Solid black line: our baseline limits derived with the 
    $V_{\rm max}$ determination of the J-factor integrated within the radius of $0.1^\circ$.
    Gray band: 95\% containment of the upper limit due to statistical fluctuations of the distribution of sources, the effect of masking the 4FGL-DR4 sources is added to the upper values of the band (appendix~\ref{app:sim-ps}).
    Solid and dashed purple lines: the limits derived by Calore et al. (figure 6 left panel%
    )~\cite{2019Galax...7...90C}. 
    Dashed orange line: limits derived by Coronado-Bl{\'a}zquez et al. (figure 12 upper panel%
    )~\cite{2019JCAP...11..045C}. 
    Solid green line: the limits derived by Butter et al. (figure 11,  the 0.8 threshold limit for $|b| > 10^\circ$%
    )~\cite{2023JCAP...07..033B}.
    Green and yellow bands with dashed gray line: limits derived for dwarf spheroidal Galaxies at 68\% and 95\% confidence level by McDaniel et al. (figure 6 top left panel)~\cite{2024PhRvD.109f3024M}.
    Dotted black line: the expectation for the WIMP DM thermal relic cross-section from ref.~\cite{Steigman_2012}.
    }
    \label{fig:bounds-bb}
\end{figure}

\begin{figure}
    \centering
    \includegraphics[width=0.80\linewidth]{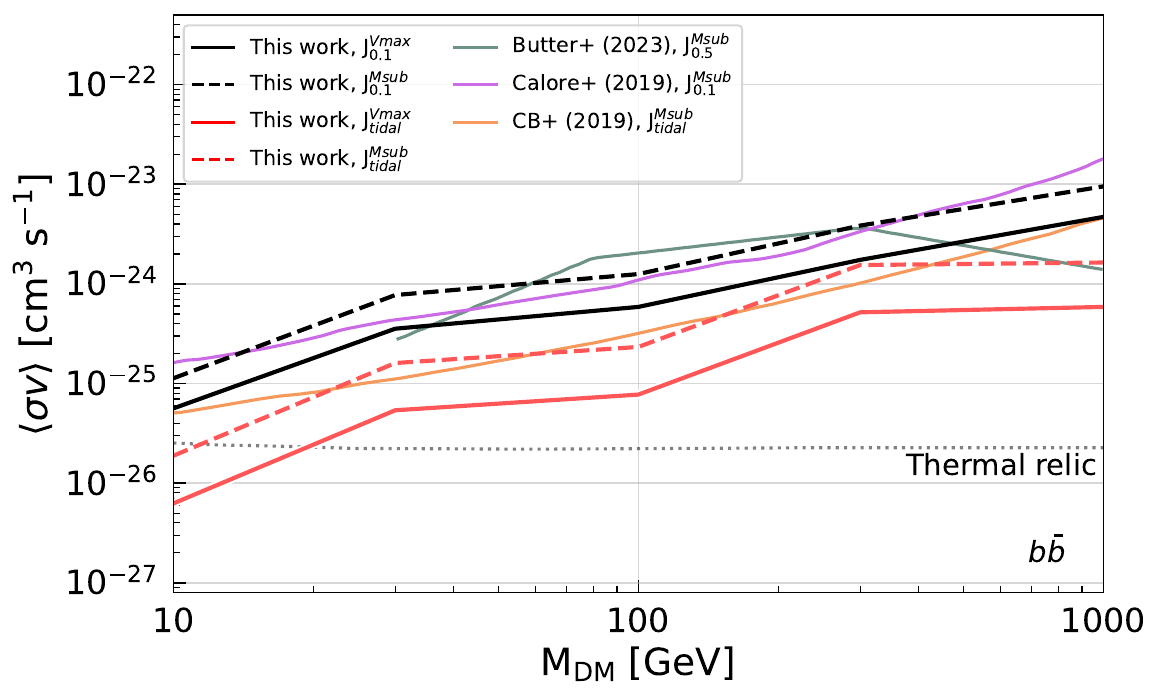}
    \caption{
    Comparison of the 95\% upper bounds on $\sigmav$ derived in this work for different choices of the J-factor.
    Solid (dashed) black line: the limit derived with the 
    $V_{\rm max}$ ($M_{\rm sub}$) determination of the J-factor integrated within the radius of $0.1^\circ$.
    Solid (dashed) red line: the limit derived with the 
    $V_{\rm max}$ ($M_{\rm sub}$) determination of the J-factor integrated within the tidal radius. 
    The uncertainty bands for the different J-factor models are proportional to the uncertainty band for our baseline model, $J_{0.1}^{Vmax}$ (Figure~\ref{fig:bounds-bb}), with the same scaling factors as the ratios of the corresponding upper limits. These uncertainty bands are not plotted for clarity of presentation.
    }
    \label{fig:bounds-bb-J}
\end{figure}

We show the upper bounds on $\sigmav$ in figure~\ref{fig:bounds-bb}.
The uncertainty band for the annihilation cross-section bound {is derived with} a bootstrap strategy. 
We identify the best-fit model for each DM mass and  use the probability density function for that model (cf. eq.~(\ref{eq:mixture})) to sample 100 new datasets via Hybrid (a.k.a. ``Hamilton'') Monte Carlo (HMC) \cite{2017arXiv170102434B} and the NUTS algorithm \cite{2011arXiv1111.4246H}. 
{For each simulated dataset, we derive the bounds on $\langle \sigma v \rangle$ and determine the $95\%$ containment interval for the resulting distribution of values. We further adjust the upper limit of this interval to conservatively account for potential systematic uncertainties associated with the presence of 4FGL sources, as detailed in appendix~\ref{app:sim-ps}. This uncertainty range is represented by the gray shaded band in figure~\ref{fig:bounds-bb}.}

The dashed gray line shows a bound derived using dwarf spheroidal galaxies~\cite{2024PhRvD.109f3024M} with 68\% and 95\% confidence bands {shown by the green and yellow areas respectively.}
We note that the constraints derived in this work are an order of magnitude less stringent than the limits derived using dwarf spheroidal galaxies (even if we consider J-factors integrated to tidal radius for the subhalos shown in figure~\ref{fig:bounds-bb-J}).
Provided that the values of J-factors $\lesssim 10^{20}\, {\rm GeV^2 cm^{-5}} $ for the brightest dwarf galaxies \citep{2024PhRvD.109f3024M} are comparable to the brightest subhalos (cf. figure~\ref{fig:Jfactors}), the main reason for the less stringent constraints is that the positions and J-factors of the dwarf galaxies are known (although with some uncertainties), while the subhalos can be at any position in the sky.

In figure~\ref{fig:bounds-bb-J}, we display a comparison of the upper bounds for different choices of the J-factor.
In particular, the limits for the J-factor integrated within $0.1^\circ$ for the $V_{\rm max}$ and $M_{\rm sub}$ J-factor
determination methods are shown by the black solid and dashed lines respectively. 
The limits corresponding to the J-factor integrated up to the tidal radius for 
the $V_{\rm max}$ and $M_{\rm sub}$ J-factor determination methods are shown by the red solid and dashed lines, respectively.

For comparison, we show the limits on $\sigmav$ obtained assuming 16 DM subhalos in the 3FGL catalog~\cite{2019JCAP...07..020C}.
The limits for J-factor integration up
to $0.1^\circ$ radius~\cite{2019Galax...7...90C} (purple solid line) are consistent with the corresponding
$J_{\rm tidal}^{M_{\rm sub}}$ model (dashed red line) within a factor of 2, comparable to the uncertainty band in figure~\ref{fig:bounds-bb}.
Larger (smaller) upper limits on $\sigmav$ relative to ref.~\cite{2019Galax...7...90C} correspond to larger (smaller) expected numbers of DM subhalos in our model compared to the 16 assumed in ref.~\cite{2019Galax...7...90C}.%
\footnote{
The $\sigmav$ upper limit depends weakly on observation time.
In the regime of a few observable DM subhalos, $dN/dJ \propto J^{-n - 1}$, so the total number of observable subhalos above a flux threshold scales as $N_{\rm sub} \propto J^{-n}$,
where $n = 1$ ($n = 2$) for the tidal radius ($0.1^\circ$) J-factor integration.
Since $F \propto \sigmav J$,
$N_{\rm sub} \propto J^{-n} \propto \sigmav^n / F^n$.
The number of background astrophysical sources in the Euclidean regime scales as
$N(> F) \propto F^{-1.5}$, where the flux threshold depends on observation time as $F \propto t^{-0.5}$.
The confidence of the upper limit on $\sigmav$
can then be estimated from the signal-to-noise ratio
$SNR = N_{\rm sub} / \sqrt{N(> F)}$.
Fixing the SNR at a given confidence level, $N_{\rm sub} = SNR \times \sqrt{N(> F)} \propto F^{-0.75}$,
and thus $\sigmav^n \propto F^n N_{\rm sub} \propto F^{n - 0.75} \propto t^{-n/2 + 0.4}$.
This gives $\sigmav \propto t^{-0.1}$ and $\sigmav \propto t^{-0.3}$ for the tidal radius and $0.1^\circ$ cases, respectively. Since these dependences are very weak, we neglect them in our estimates.

}
The limits for J-factor integration
up to the tidal radius~\cite{2019JCAP...07..020C} (orange solid line) are comparable to the corresponding limits in the
$J_{\rm tidal}^{M_{\rm sub}}$ model (dashed red line).
The more stringent constraints at the lowest (10~GeV) and highest (1000~GeV) DM masses are partly due to statistical fluctuations, which yield smaller upper limits on the number of DM subhalos (cf.\ table~\ref{tab:Nsub_values_Msub}).

The solid gray line shows the limits derived in ref.~\cite{2023JCAP...07..033B} for the J-factor integrated up to $0.5^\circ$ radius.
These limits are a factor of a few less constraining than those in the $J_{\rm tidal}^{M_{\rm sub}}$ model (dashed red line).
This can be understood from the J-factor scaling: in this model $dN/dJ \propto J^{-2}$ (red solid line in figure~\ref{fig:Jfactors}), so the total number of subhalos above a threshold $J$ is $N_{\rm sub} \propto J^{-1}$ (for the $J_{0.1}^{M_{\rm sub}}$ case, $dN/dJ \propto J^{-3}$ and $N_{\rm sub} \propto J^{-2}$).
Since $F \propto \sigmav J$, the limit scales as $\sigmav \propto N_{\rm sub}$ (or $\sigmav \propto \sqrt{N_{\rm sub}}$ in the $J_{0.1}^{M_{\rm sub}}$ case).
Two factors account for the weaker constraints in ref.~\cite{2023JCAP...07..033B}: the estimated number of subhalos (the 0.8 threshold case for $|b| > 10^\circ$) is 3 to 10 times larger than in our model (cf.\ the upper bound on $N_{\rm subhalo}^{\rm total}$ in table~\ref{tab:Nsub_values_Msub}), and the integration radius of $0.5^\circ$ is smaller than the angular extent of most DM subhalos observable from Earth~\cite{2019JCAP...11..045C, 2024MNRAS.530.2496A}.
Both effects lead to weaker constraints on the annihilation cross section compared to the $J_{\rm tidal}^{M_{\rm sub}}$ case.
An exception occurs at $M_{\rm DM} = 1$~TeV, where the limit in ref.~\cite{2023JCAP...07..033B} is a factor of 2.5 more constraining than at $M_{\rm DM} = 300$~GeV, while the expected number of DM subhalos decreases by only a factor of 2.
This is surprising, since for large DM masses the limit on $\sigmav$ at a fixed number of subhalos should scale proportionally to $M_{\rm DM}$, as is approximately the case for Coronado-Bl{\'a}zquez et al.~\cite{2019JCAP...11..045C} (solid orange line) and Calore et al.~\cite{2019Galax...7...90C} (solid purple line).
More generally, in the $J_{\rm tidal}^{M_{\rm sub}}$ ($J_{0.1}^{M_{\rm sub}}$) model we expect the limit to scale as
$\sigmav \propto N_{\rm sub} M_{\rm DM}$ ($\sigmav \propto \sqrt{N_{\rm sub}} M_{\rm DM}$).
This scaling is approximately satisfied for the last two rows in table~\ref{tab:Nsub_values_Msub} (the $J_{\rm tidal}^{M_{\rm sub}}$ case):
$10/3 \cdot {6.63/31.} \approx 0.69$,
to be compared with $1.55/1.64 \approx 0.95$,
and in table~\ref{tab:Nsub_values} (the $J_{0.1}^{M_{\rm sub}}$ case):
$10/3 \cdot \sqrt{8.61/21.46} \approx 2.1$,
to be compared with $4.7/1.7 \approx 2.8$.
Applying this scaling to estimate the $\sigmav$ limit of ref.~\cite{2023JCAP...07..033B} at $M_{\rm DM} = 1$~TeV from the value at $M_{\rm DM} = 300$~GeV, the extrapolated limit would be a factor of 4 to 6 less constraining than our red dash-dotted line at $M_{\rm DM} = 1$~TeV.
For the $J_{0.1}^{M_{\rm sub}}$ case, the larger number of subhalos and the larger integration radius in ref.~\cite{2023JCAP...07..033B}
have opposite effects on the annihilation cross-section limits.
At certain DM masses these two effects compensate each other, resulting in similar limits on the annihilation cross section
at $M_{\rm DM} = 100$ or 300~GeV (gray solid line vs.\ black dashed line in Figure~\ref{fig:bounds-bb-J}).

The limits presented in figures~\ref{fig:bounds-bb} and \ref{fig:bounds-bb-J} are derived assuming the velocity-independent DM annihilation cross section.
In case of velocity-dependent annihilation cross section, e.g., due to Sommerfeld enhancement \cite{2009PhRvD..79a5014A, Piccirillo:2022qet, Lacroix:2022cjm, Facchinetti:2022jtd}
the relative contribution of small subhalos close to the Earth, which have a smaller velocity dispersion than the dwarf galaxies, will be more important.
In particular, a significant population of subhalos at distances of a few kpc with typical velocity dispersion of a few km/s is expected \cite{2024MNRAS.530.2496A}.
These subhalos can have a factor of a few larger enhancement of annihilation cross section compared to the dwarf galaxies, where the typical velocity dispersion has a peak in the distribution around 10 km/s (e.g.,~\cite{2018JCAP...04..035L}).
The actual enhancement of the limits from subhalos compared to the dwarf galaxies depends on the details of the model (e.g., there should be no saturation of enhancement down to velocities of a few km/s), on the integral of the velocity dispersion for the subhalos, and on the relative contribution of small subhalos in the vicinity of the Earth and large distant subhalos.
This detailed analysis is differed to a future work.

\section{Discussion and conclusions}
\label{sec:conclusions}
In this work, we have searched for DM annihilating subhalos among the unassociated \Fermi-LAT sources. We developed the first fully probabilistic model of gamma-ray sources, incorporating both known astrophysical source classes and a potential new population of gamma-ray sources modeled as DM subhalos annihilating into Standard Model particles. 

We characterize gamma-ray sources using their  energy spectra fitted by the log-parabola function.
We take the parameters $\bf x$ of the log-parabola fit as the input features. 
We then
construct a probabilistic model $p({\bf x})$ for the unassociated sources as a weighted mixture of three source classes: Galactic and extragalactic astrophysical sources plus the hypothetical DM subhalos. The Galactic and extragalactic components encompass various source types classified by the \Fermi-LAT collaboration. We analyze data shifts between associated and unassociated sources within a general framework of quantification learning, considering two effects: (1) a covariate shift affecting all source classes in the same way, and (2) a prior probability shift, which allows for changes in class prevalence, including the possible presence of a new source class among the unassociated sources.

The astrophysical components of the mixture model are the conditional PDFs $p(\mathbf{x}|k)$ for the two astrophysical classes $k$, Galactic and extragalactic.
They are built from the corresponding empirical distributions of the associated sources, modified by an optimizable covariate-shift function. The DM subhalo component of the mixture is determined by Monte Carlo simulations based on a set of DM parameters (mass, annihilation channel, and cross section), including the distribution of J-factors coming from N-body simulations. The global likelihood of the model is obtained by multiplying the PDFs $p({\bf x}_i)$ of the unassociated sources $i$ with ${\bf x}_i$ inside the range of parameters that we consider. 

By optimizing the model likelihood, we extract the best-fit parameters, including $\sigmav$, for various DM masses in the $b\bar{b}$ annihilation channel.
No significant excess corresponding to DM subhalos is observed beyond Galactic and extragalactic sources.
Consequently, we derive 95\% upper bounds on the DM annihilation cross section. To our knowledge, this is the first time an upper bound on DM annihilation has been obtained by constraining the contribution of DM subhalos to unassociated \Fermi-LAT sources using a maximum likelihood approach.
Our constraints are competitive with previous limits derived from DM subhalo searches using classify-and-count methods.
With respect to the latter, our approach, based on quantification learning, offers several advantages :
\begin{enumerate}[label=(\roman*)]
    \item It accounts for differences in distributions (data shifts) between associated and unassociated source populations.
    \item It determines the potential contribution of a new hypothetical source population, e.g., the DM annihilating subhalos, without imposing ad hoc assumptions on its fraction relative to known sources. This is a key advantage over traditional supervised machine learning classification.
    \item The probabilistic model enables one to generate mock data (samples of unassociated sources), which we use to validate our approach and estimate the uncertainties of the upper bounds on the DM annihilation cross section. 
    \item The model provides both the PDF $p(\mathbf{x}|k)$ of observed source features $\mathbf{x}$ given a source class $k$ (either astrophysical or DM), and the posterior probability $p(k|\mathbf{x})$ of a source belonging to a given class based on observed features.
\end{enumerate}
Thus, our approach allows one not only to determine
the conditional probabilities $p(k|\mathbf{x})$, which is the main objective of ML classification models, but also to estimate the overall likelihood of the model given the distribution of unassociated sources. This likelihood serves as a tool to search for, or constrain, a new class of sources absent from the training data, i.e., the associated \Fermi-LAT sources.

This work bridges the gap between ML approaches for characterizing unassociated sources traditionally focused on classification and the standard astrophysical analysis using maximum likelihood models. As a result, we construct a model capable of both classifying sources and constraining a new source population. 
Our approach is broadly applicable to searches of anomalies in distributions for datasets exhibiting both covariate and prior shifts, where the target data is unlabeled.

\acknowledgments
AA and BZ acknowledge the support from Generalitat Valenciana through the “GenT program”, ref.: CIDEGENT/2020/055.
The work of DM is supported by the DFG grant MA 8279/3-1. 
The work of VG was supported by PID2022-139841NB-I00 and PID2021-125331NB-I00. 
The work of MASC was supported by the grants PID2021-125331NB-I00 and CEX2020-001007-S, both funded by MCIN/AEI/10.13039/ \linebreak 501100011033
and by ``ERDF A way of making Europe''. MASC also acknowledges the MultiDark Network, ref. RED2022-134411-T. 
The authors gratefully acknowledge the computer resources  of ARTEMISA and the technical support provided by the Instituto de Fisica Corpuscular, IFIC(CSIC-UV). ARTEMISA was created thanks to the European Union through the 2014-2020 ERDF Operative Programme of Comunitat Valenciana, project IDIFEDER/2018/048, and nowadays it is supported by the grants
EUR2022-134028, and ASFAE/2022/024. 
We acknowledge the use of several statistical and optimization tools, including Jax \cite{jax}, Tensorflow Probability \cite{tensorflow2015-whitepaper}, Optim.jl \cite{mogensen2018optim}, Healpy \cite{healpy}, and Scipy \cite{scipy}.

\bibliography{DM_halos}%

\pagebreak

\appendix

\section{Simulation of gamma-ray signals from dark matter subhalos}
\label{app:sim-ps}

In this appendix we provide details about the simulation of gamma-ray signal from DM subhalos.
Eq.~(\ref{eq:pDM}) describes the probability of observing a DM subhalo 
given a DM model.
This equation includes three terms that need to be estimated through simulations: 
the number density $n(\phi_{\rm DM})$ of DM subhalos as a function of expected flux (section~\ref{sec:DM_model}), the probability $p_{\rm DM}(\alpha,\beta|\phi_{\rm DM};\btheta_{\rm DM})$  of observing a source with the log-parabola spectrum parameters $(\alpha,\,\beta)$ given a DM model, which we obtain by computing and fitting simulations of DM point sources, and the efficiency $\varepsilon_{\rm eff}(\phi_{\rm DM})$ of the detection of a source, which we obtain from the same set of simulations by identifying the fraction of sources which can be detected by \Fermi-LAT.  
We now describe how the DM distribution conditioned on the gamma-ray flux is derived and how we estimate the detection efficiency for DM subhalos.

We compute the DM distribution  conditioned on the gamma-ray flux, through realistic photon-counts simulations of DM point sources. 
We decompose the gamma-ray sky into two  components: a diffuse emission component (constituted by the diffuse Galactic emission and an isotropic background) and the signal from the DM point sources. 
We model the diffuse Galactic emission using the \verb|gll_iem_v7| template provided by the \Fermi-LAT collaboration as well as the corresponding isotropic background model.
We integrate eq.~(\ref{eq:DM_spec}) in 16 logarithmic energy bins between 100 MeV and 1 TeV to obtain the theoretical energy spectrum in units of the photon flux per energy bin.
We specify the spectral energy distribution (SED), $dN_\gamma/dE$, using the recent estimations from CosmiXs \cite{2024JCAP...03..035A}.

Furthermore, we notice that the J-factor and annihilation cross-section term factor out of the integral into an overall constant that ultimately defines the total flux normalization: 
\begin{equation}
   \Phi_i = A \int_{E_{min, i}}^{E_{max,i} }\frac{dN_\gamma}{dE}dE \,.
\end{equation}

For this reason, once we fix the SED for a specific DM annihilation channel and mass, the energy spectrum $\phi(E)$ is defined except for an energy-independent normalization term. 
By fixing the overall flux normalization, we obtain the expected gamma-ray flux due to a specific DM source. 
Experimental uncertainties in the reconstruction of energy and direction of gamma rays induce a spread in the distribution of photons associated with a point source.
We simulate sources by convolving with the \Fermi-LAT point spread function (PSF) and exposure
using {\Fermi} Science Tools (fermitools) version 2.2.0\footnote{\url{https://github.com/fermi-lat/Fermitools-conda/wiki}}.

In table \ref{tab:settings} we summarize the analysis settings, which are similar to the standard event selection cuts used in the construction of gamma-ray source catalogs. 
Likewise, we convolve the diffuse Galactic foreground and isotropic emission with the PSF and exposure to obtain the diffuse foreground emission.
For each simulation, we add the emission from 10,000 DM sources randomly and isotropically distributed in the sky together with the foreground emission. We chose this number of sources to roughly match the number of resolved sources in the 4FGL catalog and to have sufficient statistics for the DM sources, while avoiding excessively demanding simulations. We thus obtain a map for the expected photon counts due to a DM population with a specific mass, annihilation channel, and a common flux.

\begin{table}[t]
\centering
\begin{tabular}{ | l| l|  }
\hline
Healpix order & 8\\ 
Weeks & $9-795$\\ 
Emin & 100 MeV\\
Emax & 1 TeV\\
Instrument Response Functions (IRFs) & \texttt{P8R3\_SOURCE\_V3} \\
EVCLASS & 128 (Source)\\ 
EVTYPE & 3 (Front+Back)\\
ZMAX & $90^\circ$ under 1 GeV, $105^\circ$ otherwise\\
Galactic foreground model & \verb|gll_iem_v7.fits| \\
\hline
\end{tabular}
\caption{\texttt{Fermi Science Tools} settings used for the 15-year data set analysis.}
\label{tab:settings}
\end{table}

We employ each simulation (associated with a specific total gamma-ray flux for the DM source) to estimate $p_{\rm DM}(\alpha,\beta|\phi_{\rm DM};\btheta_{\rm DM})$ as follows.
First, we perform a Poisson realization of the expected photon counts map.
Then we proceed with fitting the energy spectrum of each source with a log parabola function (with a fixed spatial position), yielding the DM distribution conditioned on the flux.
Finally, we estimate the detection efficiency $\varepsilon_{\rm eff}(\phi_{\rm DM})$ by studying the TS of the DM sources for a given map. We define the TS of a source by comparing the overall likelihood with and without the target source. We define the TS as:
\begin{equation}
TS = -2\ln\left(\frac{\mathcal{L}}{\mathcal{L}_0}\right),
\end{equation}
where $\mathcal{L}$ is the likelihood of the best fit model including the source, and $\mathcal{L}_0$ is the likelihood of the model without the source. 
If the TS of a source is larger than 25 we include that source to a database of resolved sources, for which the SED fit parameters are reliable and can be used for the estimation of $p_{\rm DM}(\alpha,\beta|\phi_{\rm DM};\btheta_{\rm DM})$. Else, we discard the source as a low TS source which would not be measured by the {\Fermi} LAT.

The detection efficiency $\varepsilon_{\rm eff}(\phi_{\rm DM})$ of the {\Fermi} LAT to a DM source with a specific flux and SED is then defined as the ratio between the number of detected sources and the overall number of sources included in the simulation: 
\be
\varepsilon_{\rm eff} = \frac{N_{\rm sources}(TS>25)}{N_{\rm sources}^{\rm tot}}.
\ee

\begin{figure}
    \centering
    \includegraphics[width=0.5\linewidth]{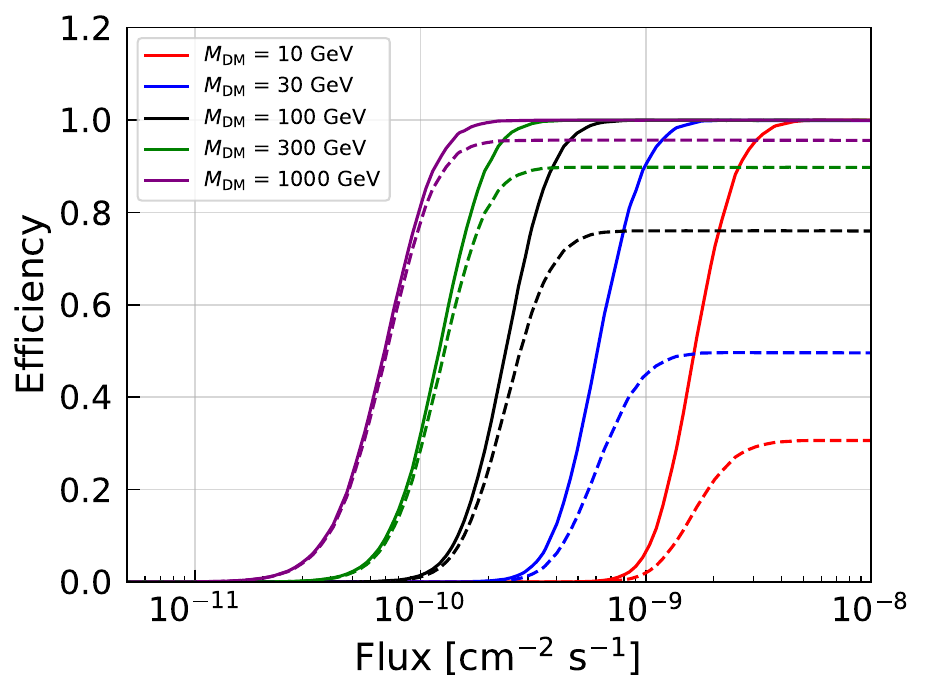}
    \caption{Detection efficiency for the $b\overline{b}$ annihilation channel simulations. {Solid lines denote the baseline efficiency used in the analysis, derived from simulations that disregard known 4FGL sources. Dashed lines illustrate a conservatively corrected efficiency, representing a worst-case scenario where a disk around each 4FGL source is masked to estimate the maximum potential degradation caused by their inclusion.}}

    \label{fig:efficiency}
\end{figure}

In order to avoid biasing the estimation of the DM distribution due to a low number of detectable sources or due to a very specific realization of the sky, we perform additional simulations of the gamma-ray sky until we detect a total of at least 20,000 DM sources or until 1,000 simulations have been performed. 
Once we have built a dataset of measurements for the energy spectrum of DM sources, we estimate the probability density function $p_{DM}(\alpha, \beta|\phi_{DM})$ by employing KDE, in analogy with what we did in section~\ref{sec:stat-model} for the astrophysical sources.

Due to the high number of DM sources, and the large number of simulations, we cannot employ the full detection and characterization pipeline which is usually adopted by the \Fermi-LAT collaboration.
We exploit the additional information available to us from simulations to perform a simplified analysis pipeline. 
The true energy spectrum of each DM source is fixed (and known) in our simulations, and so are the foreground components. The positions of the DM sources are known as well, and the sources are all point-like sources. For these reasons, we make the following assumptions in our analysis:
\begin{itemize}
    \item We assume that the energy spectrum of each background source (i.e. any source different from the one we are fitting) is known and fixed to the theoretical one (while in general they are jointly optimized together with the target of the fit);
    \item We fix the diffuse Galactic and the isotropic background components;
    \item We assume that the localization of the source is known (which in this case is true, but not in the case of source detection on real data);
    \item We assume that there are no low TS background sources, as we know exactly what we put inside our simulation.
    \item {We assume that contributions from known 4FGL sources are negligible when estimating the DM distribution. While these bright sources can reduce detection efficiency and should ideally be modeled jointly, we estimate the maximum degradation by masking a disk around each 4FGL source. 
    This mask corresponds to the 68\% containment angle of the PSF at the peak energy ($E_{peak}$) of the DM spectrum for each considered mass. Because the peak of the DM energy spectrum increases with the DM mass, the detection efficiency is affected more significantly at lower DM masses, as displayed in figure~\ref{fig:efficiency}. Specifically, for $\mdm$ $\geq$100 GeV, the narrower PSF results in a masked sky fraction of 20\% or less, deteriorating the bounds by approximately 15\%. 
    Conversely, for lower masses ($\mdm = 10 - 30$ GeV), the broader PSF yields a masked fraction of $50 - 70$\%, leading to a $35 - 60$\% degradation of the bounds. 
    Except for the $\mdm$ = 10 GeV case, this degradation remains within the bootstrap uncertainty band derived for the unmasked analysis. Nonetheless, we conservatively adjust the 95\% upper confidence level in figure~\ref{fig:bounds-bb} to account for these systematic effects.}
\end{itemize}

By making these assumptions, we fit each DM source individually, regardless of the others.

To perform a maximum likelihood estimation fit of the DM energy spectrum for each source, we define the likelihood as follows:
\begin{equation}
    \mathcal{L} = \prod_{e=1}^{N_{bin}} \prod_{j \, \in \, {\rm disc}_e}  
    \mathcal{P}(k_{e,j}, \lambda_{e,j}(\bx)).
\end{equation}
Specifically, for each energy bin, we identify a disc around the source corresponding to the 95\% containment angle of the PSF with a minimal angle of $0.5^\circ$ and a maximal angle of $5^\circ$.
This ensures high enough statistics for the high-energy bins, and reduces the background noise for the low-energy bins, while leaving the intermediate-energy bins unaffected.

For each pixel of the disc, we compare the photon counts $k_{ij}$ with the estimated photon counts $\lambda_{ij}(\bx)$ using the Poisson likelihood. The estimated photon counts are a sum of the Galactic foreground in the disc, the isotropic background, the emission of the other background (fixed) DM sources and the photon counts coming from the target source:
\begin{equation}
    \lambda(\bx) = \lambda_{gal} + \lambda_{iso} + \lambda_{S \neq S_i} + \lambda_{S_i}(\bx).
\end{equation}
Specifically, for a source $i$ and pixels $j$:
\begin{equation}
    \lambda_{S_i, e,j} = \left[\int_{E_{min}^e}^{E_{max}^e} \phi_{0,i}\frac{E}{E_0}^{-\alpha_i - \beta_i\log(E/E_0)} dE\right] \cdot {\rm PSF_{i,j}^e} \cdot {\rm exposure_{j}^e} \cdot \frac{4\pi}{12 N_{\rm side}^2}.
    \label{eq:lambda}
\end{equation}
We repeat the fitting procedure for each DM source, at a fixed total flux, and we build the DM distribution $p_{DM}(\alpha, \beta|\phi)\equiv p_{DM}^{KDE}(\alpha, \beta|\phi)$ through the KDE.

\section{Details of model optimization with the EM algorithm}
\label{app:EM}

Maximum likelihood estimation of the parameters of a mixture model is a challenging task due to the interplay between the mixture weights and the distribution parameters. 
A standard iterative algorithm for this task is the  Expectation Maximization (EM) algorithm \cite{bishop, hastie2009elements}. 

We introduce latent binary vectors ${\bf z}_i$, corresponding to each data point $\bx_i$, where dim(${\bf z}_i)=K$, where $K$ is the number of mixture components. These latent variables encode the membership of point $\bx_i$ in the $k$-th component of the mixture, such that all elements of ${\bf z}_i$ are zero, except for $z_{ik}=1$. In other words, ${\bf z}_i$ is a categorical random variable represented with the one-hot-encoding scheme. The coefficients $\pi_k$ of the likelihood in eq.~(\ref{eq:mixture}) are interpreted as the parameters determining the corresponding categorical {\it prior} distribution of ${\bf z}_i$, i.e. $p(z_{ik}=1)=\pi_k~\forall i$, such that  
\begin{equation}
 p({\bf z}_i) = \prod_k^K \pi_k^{z_{ik}} ~.  
\end{equation}
With this interpretation for ${\bf z}_i$, we can identify $ p(\bx_i | z_{ik}=1) = p_k(\bx_i|\btheta_k)$, such that
\begin{equation}
p(\bx_i|{\bf z}_i) = \prod_k^K p_k(\bx_i|\btheta_k)^{z_{ik}}~.
\end{equation}
We then have all the elements to compute the {\it posterior} distribution of ${\bf z}_i$ as:
\begin{equation}
p(z_{ik}=1 | \bx_i) = \frac{p_k(\bx_i|\btheta_k) \pi_k}
{\sum_j^K p_j(\bx_i|\btheta_j) \pi_j} \equiv \gamma(z_{ik})~.
\label{gamma}
\end{equation}

In order to find the optimum parameters $\{\btheta_k\}$ and weights $\boldsymbol{\pi}$, the idea is to consider the joint likelihood $p(\bx_i, {\bf z}_i)$ 
for the whole dataset of points $\bx_i$ with associated latent variables ${\bf z}_i$:
\begin{equation}
\label{p(X,Z)}
p({\bf X,Z}|\{\btheta_k\},\boldsymbol{\pi})
=\prod_i^N \prod_k^K [\pi_k p_k(\bx_i|\btheta_k)]^{z_{ik}}~.
\end{equation}
In spite of this apparent complication of introducing extra variables in the problem, i.e., the latent variables, the likelihood in eq.~(\ref{p(X,Z)}) is arguably easier to optimize than the original likelihood $p(\bx|\{\btheta_k\},\boldsymbol{\pi}) = \prod_i p(\bx_i)$, where each term in the product is built as in eq.~(\ref{eq:mixture}). One of the reasons for this is that, the log-likelihood of eq.~(\ref{eq:mixture}) is a log of a sum, which is computationally challenging to optimize, especially in the large $K$ regime, while the log of the likelihood in eq.~\ref{p(X,Z)} is the sum of logs, i.e., $\log p({\bf X,Z}|\{\btheta_k\},\boldsymbol{\pi})$ is very similar to the cross-entropy cost function, which is easily optimizable.  

An obvious problem is that in eq.~(\ref{p(X,Z)}) we do not know the $z_{ik}$ variables. The EM algorithm considers instead the expectation value of this likelihood (or more concretely, of the log-likelihood), under the posterior distribution of ${\bf Z}$, i.e.
\begin{equation}
\label{E_log}
\mathbb{E}_{\bf Z} 
[\log p({\bf X,Z}|\{\btheta_k\},\boldsymbol{\pi}) ] =
\sum_{\bf Z} 
p({\bf Z}|{\bf X},\{\btheta_k\},\boldsymbol{\pi})~
\ln p({\bf X,Z}|\{\btheta_k\},\boldsymbol{\pi})~.
\end{equation}

Eq.~(\ref{E_log}) is maximized according to the following iterative procedure:
\begin{enumerate}
    \item Initialize $\{\btheta_k\}=\{\btheta_k\}_{\rm old}$, and $\boldsymbol{\pi} = \boldsymbol{\pi}_{\rm old}$;
    \item Evaluate $p({\bf Z}|{\bf X},\{\btheta_k\}_{\rm old},\boldsymbol{\pi}_{\rm old})$;
    \item Plug the posterior obtained in step 2 into expr.(\ref{E_log}), such that now the only term containing the optimizable parameters $\{\btheta_k\},\boldsymbol{\pi}$ is the log-likelihood. So we get the expression:
    \begin{equation}
      Q(\{\btheta_k\},\boldsymbol{\pi}) \equiv
     \sum_{\bf Z} 
    p({\bf Z}|\bx,\{\btheta_k\}_{\rm old},\boldsymbol{\pi}_{\rm old})~
\ln p({\bf X,Z}|\{\btheta_k\},\boldsymbol{\pi})~.  
    \end{equation}
    \item Update the parameters as:
    \begin{equation}
      \{\btheta_k\}_{\rm new},\boldsymbol{\pi}_{\rm new} =
      \underset{\{\btheta_k\},\boldsymbol{\pi}}
      {\rm argmax}~Q(\{\btheta_k\},\boldsymbol{\pi})
    \end{equation}
    \item Repeat above steps until convergence
\end{enumerate}
Note that in step 4 above, the optimization of $\boldsymbol{\pi}$ is analytical, i.e., using a Lagrange-multiplier optimization with the condition $\sum_j \pi_j = 1$, we obtain:
\begin{equation}
    (\pi_k)_{\rm new} = \frac{1}{N}\sum_i^N \gamma(z_{ik})~,
\end{equation}
where $\gamma(z_{ik})$ is given by eq.~(\ref{gamma}). The optimization of $\{\btheta_k\}$ is in general not analytical.

\section{Consistency checks of the model}
\label{app:MC}

In this appendix we perform several checks of our model.
At first, we investigate the relative importance of prior and covariate shifts by comparing models with only prior or only covariate shift to the full model that contains both prior and covariate shifts. 
We then proceed to perform consistency checks. We start by generating distributions of sources in parameter space from our best-fit model (including only astrophysical components) and fit the generated data with the same model.
The goal is to determine a distribution of $\log\mathcal{L}$ values due to statistical fluctuations (since the data was generated from this model) and to compare to the value of $\log\mathcal{L}$ for the actual data to check if the model without a DM contribution is generally a good fit to the data.
As a last consistency check, we perform injection tests, where we add a DM subhalo component to the astro-only model to check if we can recover the injected signal.

\subsection{Importance of covariate and prior shifts}
\label{app:prior_vs_cov}

In order to investigate the importance of the prior and covariate shifts, 
we first compare the performance of four models in eq. (\ref{eq:mixture}) without the DM component: a model without either prior or covariate shifts (i.e., no free parameters) determined from associated sources, denoted as ``$as$'', models with only prior or covariate shifts, denoted as ``$p$'' or ``$c$'' respectively, and a model with both prior and covariate shifts, denoted as ``$p+c$''. 

The model without either prior or covariate shifts is constructed from the catalog of associated sources following sections \ref{sec:cov_prior} and \ref{sec:stat-model}, where the prior probability of a class is given by the relative class prevalence $\pi_k$ in the 4FGL-DR4 catalog, while the PDFs of the Galactic and extragalactic components are derived from the corresponding associated sources.
The model with either prior or covariate shift follows eq. (\ref{eq:mixture}), 
where we fix the covariate shift modulation to be equal to 1 (i.e., $\tilde{C}(\bx; \btheta_{\rm cov}) = 1$), or fix the prior probabilities to the corresponding values for associated sources.

\begin{table*}[t]
\centering
\resizebox{0.95\textwidth}{!}{
\begin{tabular}{|lcccc|ccc|}
\toprule
model & $2\Delta \log \LL_{as,x}$ & $\Delta{\rm dof}_{as,x}$ & $N\sigma_{as,x}$ & ${\rm AIC}_{as,x}$ & $2\Delta \log \LL_{x,p+c}$ & $\Delta{\rm dof}_{x,p+c}$ & $N\sigma_{x,p+c}$\\
\midrule
associated          & 0       & 0  & 0.0  & -178 &  64.7 & 7 & 6.7\\
prior               & 49.5    & 1  & 7.0  & -225 & 14.2 & 6 & 2.2\\
covariate           & 33.4    & 6  & 4.5  & -199 & 30.3 & 1 & 5.5 \\
prior + covariate   & 64.7    & 7  & 6.7  & {-227} & 0.0  & 0 & 0.0 \\
\bottomrule
\end{tabular}
}
\caption{Comparison of models of unassociated sources with Galactic and extragalactic components. ``as'' denotes the model derived from distributions of associated sources applied to the unassociated sources (without free parameters). ``$p+c$'' is the full covariate + prior shift model. 
Subscipt ``$x$'' denotes the model specified in the first column of the table.
For example, the value of $2\Delta \log \LL_{as,x} = 33.4$ in the fourth row and second column corresponds to the difference between the model derived for associated sources and the model with only covariate shift, i.e., to $x = c$ (covariate) in this case.
}
\label{tab:models}
\end{table*}

We compare the performance of the four models evaluated on the unassociated sources in the ROI for $E_0$ = 1 GeV in table \ref{tab:models}.
In column 2, we show the difference in log likelihood between the model specified in the first column and the model based on distributions of associated sources.
We report the corresponding additional degrees of freedom in column 3.
We find that the prior shift model leads to a larger increase in likelihood for one degree of freedom (we have a constraint that the sum of priors is equal to 1: $\pi_{\rm gal} + \pi_{\rm ext} =1$) compared to a model with covariate shift only.
The corresponding significance of exclusion of the null hypothesis (the ``as'' model) is presented in column 4.
It is derived by assuming the $\chi^2$ distribution for $-2\Delta \log \mathcal{L}$ with the corresponding number of degrees of freedom from column 3.
In column 5 we show the Akaike information criterion, ${\rm AIC} = -2\log\mathcal{L} + 2 k$, where $k$ is the number of degrees of freedom.
We find that although log likelihood of the full prior $+$ covariate shift model is the largest,
both the significance of exclusion of the null hypothesis and the AIC for this model are comparable to the ones for the prior shift only model.
Thus, we conclude that the prior shift is the dominant effect compared to the covariate shift.
We further compare the models in column 1 to the full prior $+$ covariate shift model in columns 6, 7, and 8, where we show $2\Delta \log \mathcal{L}$, the additional degrees of freedom, and the significance of exclusion of the null hypothesis, which, in this case, is the model in the first column.
In particular, we find that the prior shift model can be excluded only at 2.2 sigma level in favor of the ``$p + c$'' model.

\begin{figure}[t]
    \centering
    \resizebox{0.95\textwidth}{!}{
    \includegraphics[width=0.30\linewidth]{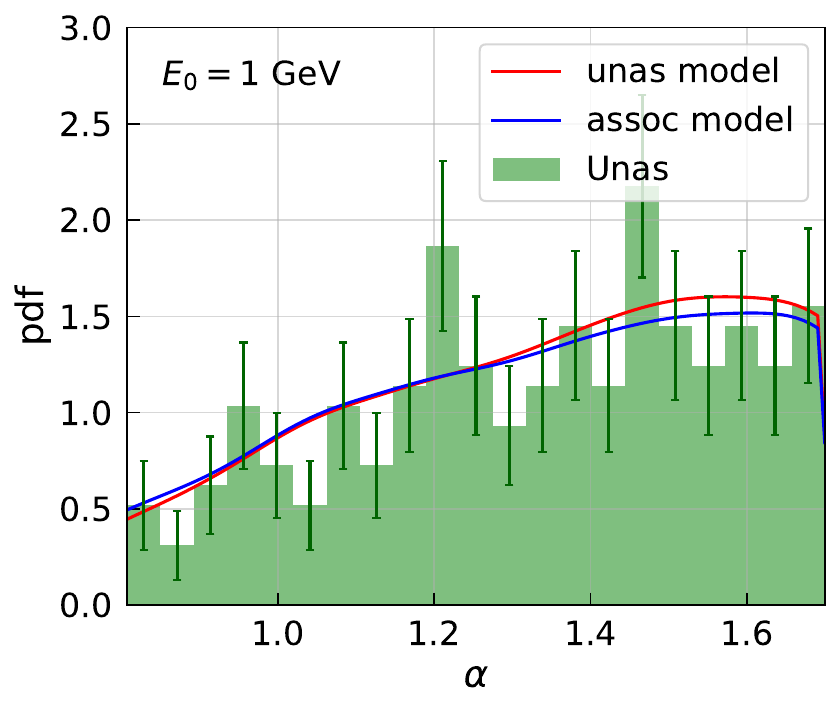}
    \includegraphics[width=0.30\linewidth]{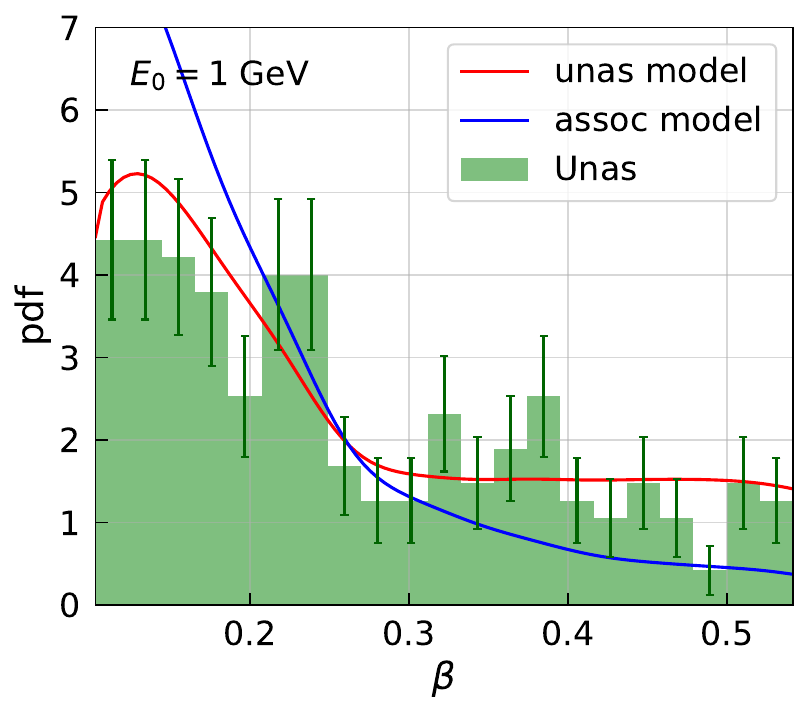}
    \includegraphics[width=0.30\linewidth]{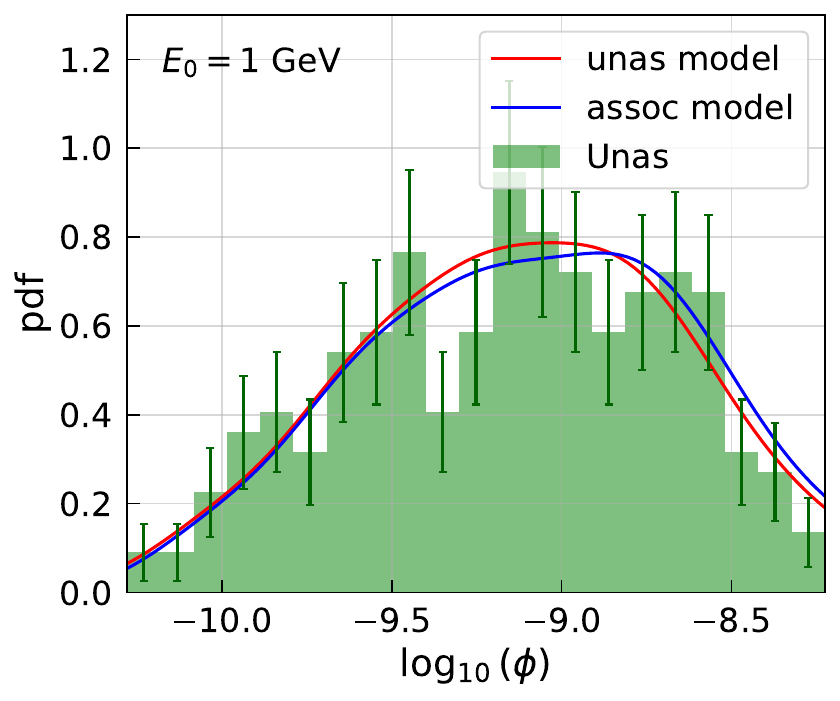}
    }
    \caption{Distributions of the $\alpha$, $\beta$ and $\log_{10}(\phi)$ parameters 
    for the unassociated sources in the 4FGL-DR4 catalog. 
    The bars show statistical uncertainty.
    The best-fit model that includes both covariate and prior shifts is shown by the red line while the model derived for associated sources is shown by the blue line. The 3D model is integrated in two variables to obtain the distribution in the third variable, shown on the plot.
    }
    \label{fig:1d-marginals}
\end{figure}

In figure~\ref{fig:1d-marginals},
we compare our models with the distributions of parameters $\alpha$, $\beta$, and $\log_{10}(\phi)$ for the unassociated sources. 
We display the best-fit ``$p+c$'' model (in red) and compare it with the prior KDE model derived for associated sources (in blue).
The ``$p+c$'' model provides a reasonable fit to the data for all the distribution parameters. On the other hand, the model derived for associated sources fails to capture the $\beta$ dependence correctly, as expected, since the distribution for unassociated sources differs considerably from the distribution of associated sources for the $\beta$ parameter (cf. figure~\ref{fig:alpha_beta_hist_E01000MeV}).

\subsection{Bootstrap MC and overall goodness of fit}
\label{app:model_goodness}

\begin{figure}[t]
    \centering
    \resizebox{0.95\textwidth}{!}{
    \includegraphics[width=0.45\linewidth]{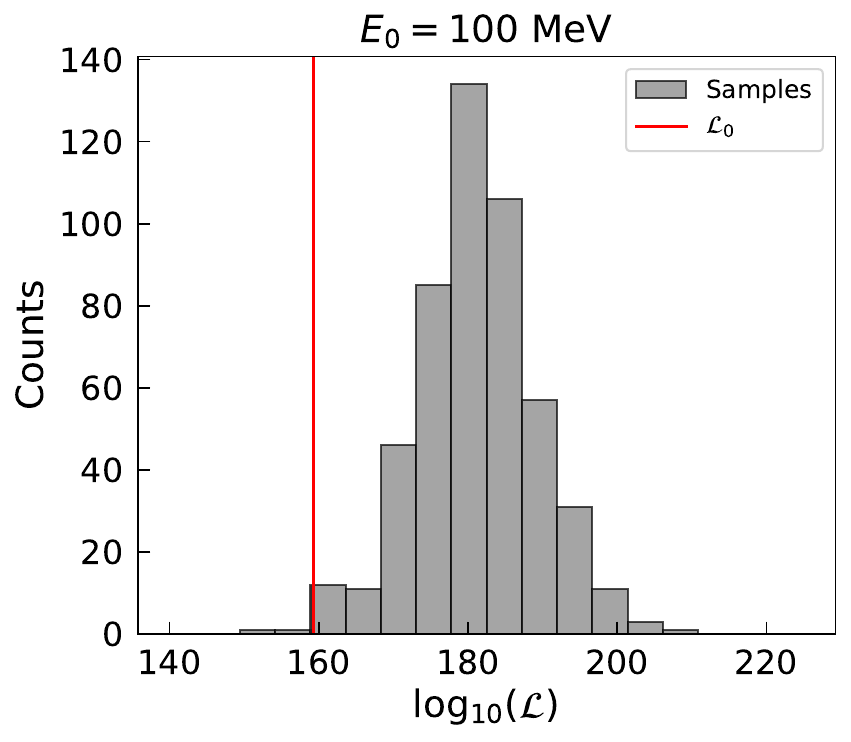}
    \includegraphics[width=0.45\linewidth]{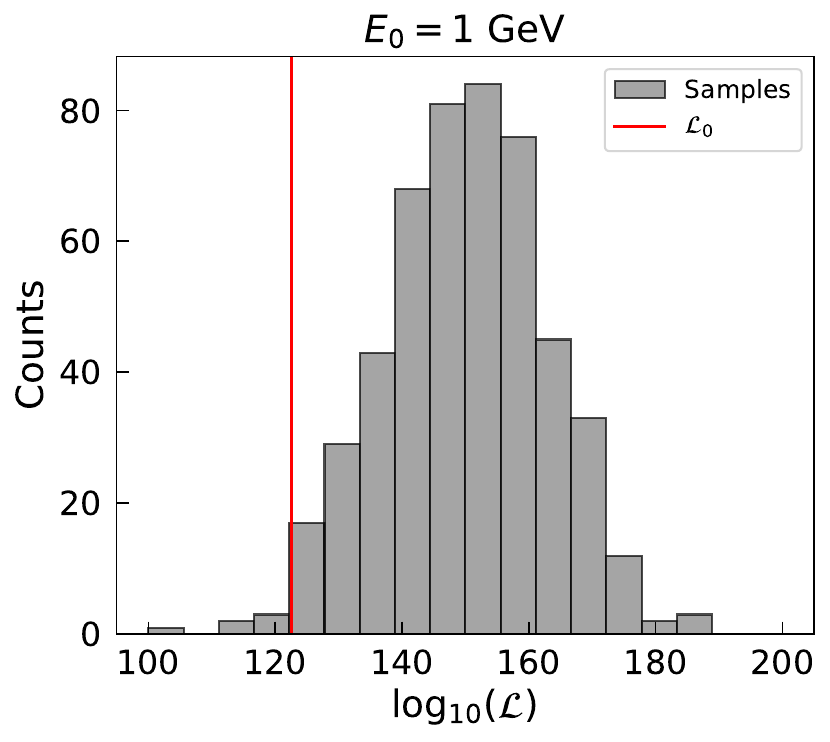}
    }
    \caption{Distribution of log-likelihood for the ``$p+c$'' model with astro-only components for 1000 datasets simulated from the best-fit ``$p+c$'' model for $\rm E_0=100 \, {\rm MeV}$ (\panel{left panel}) and $\rm E_0=1 \, {\rm GeV}$ (\panel{right panel}). The red line represents the log-likelihood of the ``$p+c$'' model with astro-only components for the unassociated 4FGL-DR4 sources.}
    \label{fig:loglike-bootstrap}
\end{figure}

We quantitatively estimate the goodness of fit of the ``$p + c$'' model in figure~\ref{fig:loglike-bootstrap}.
In this figure, we generate samples of sources in the parameter space from the best-fit ``$p + c$'' model. We then reoptimize the model for each of the samples.
The corresponding distribution of the log likelihood values has only statistical fluctuations.
The red lines on these plots show the best-fit $\log \mathcal{L}$ value for the real data.
These values are about 2 sigma smaller than the mean of the $\log \mathcal{L}$ distribution both for $E_0 = 100$~MeV (left panel) and for $E_0 = 1$~GeV (right panel).
On the one hand, this deviation shows that the model is not perfect, i.e., there are possibly some deviations of the model and the data.
On the other hand, the deviation has only about 2 sigma statistical significance.
This deviation may be connected to the slight excess observed, e.g., in the $\mdm = 30$~GeV DM annihilation (cf. the $\mdm = 30$~GeV row in Table~\ref{tab:Nsub_values}).
{Indeed, if there is an excess at about $2\sigma$ significance, which is not accounted by the model, then the model would be excluded at $2 \sigma$ significance compared to a model that can explain the data without excesses, e.g., the data generated from the model itself (cf. figure~\ref{fig:loglike-bootstrap}).}

\subsection{DM subhalo injection test}
\label{app:DM_injection}

In order to ensure our model is actually capable of reconstructing a DM signal, we inject a DM signal into simulated data and test the ability of our model to reconstruct it.
We choose a significant value of $\sigmav$ for each DM mass. For $\mdm$ = 30, 100, 300 GeV, we adopt the best fit $\sigmav$ value from table \ref{tab:Nsub_values}, while for $\mdm = 10, 1000$ GeV, since the best-fit value is zero, we inject a signal with a $\sigmav$ corresponding to the 3$\sigma$ significance (which should be clearly distinguishable from the null hypothesis). 

\begin{figure}[t]
    \centering
    
    \includegraphics[width=0.45\linewidth]{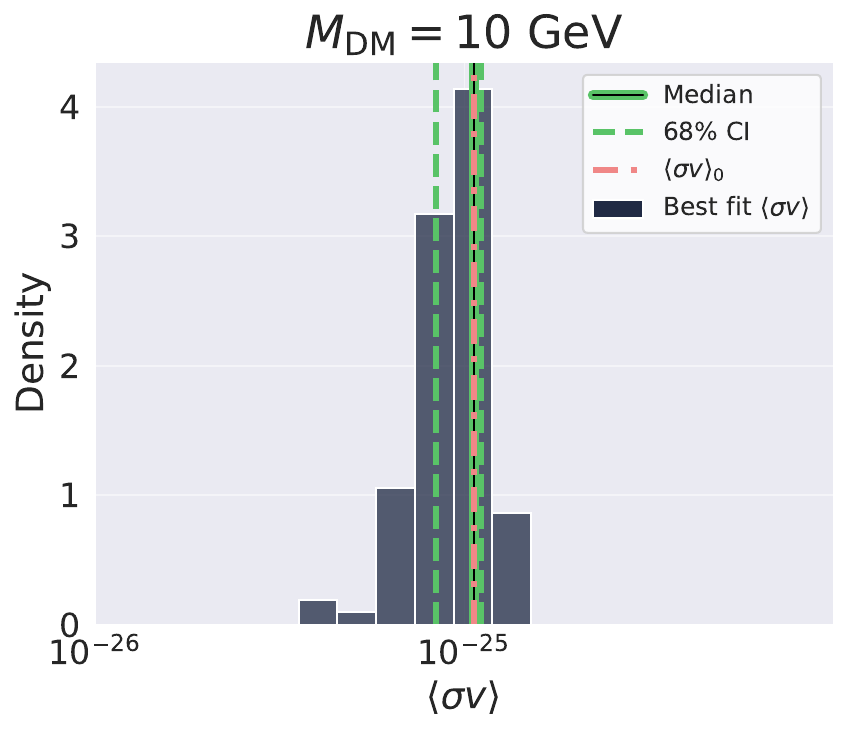}
    \includegraphics[width=0.45\linewidth]{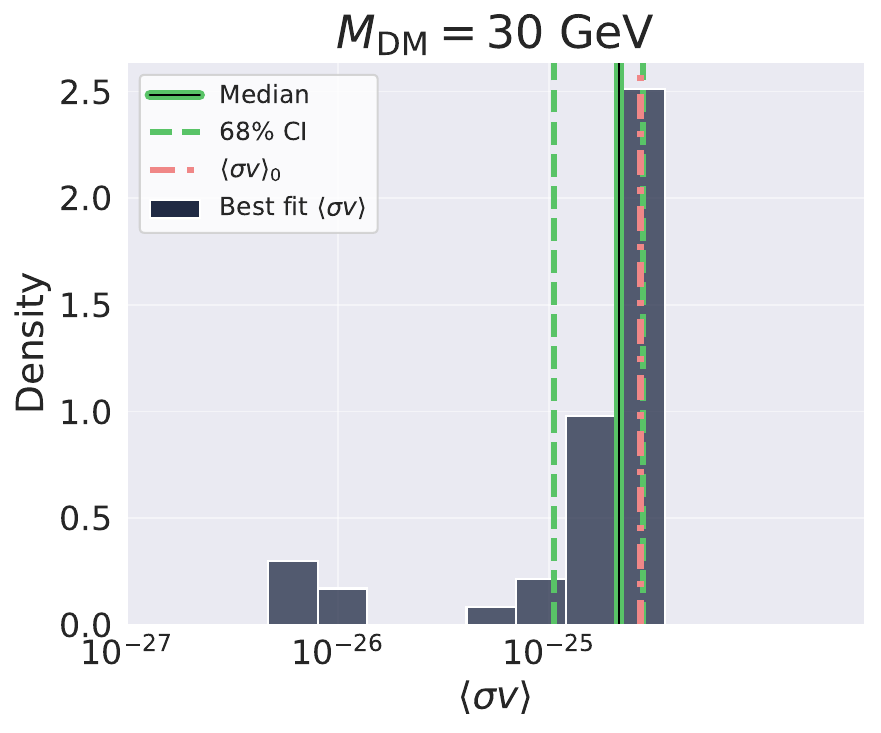}
    \includegraphics[width=0.45\linewidth]{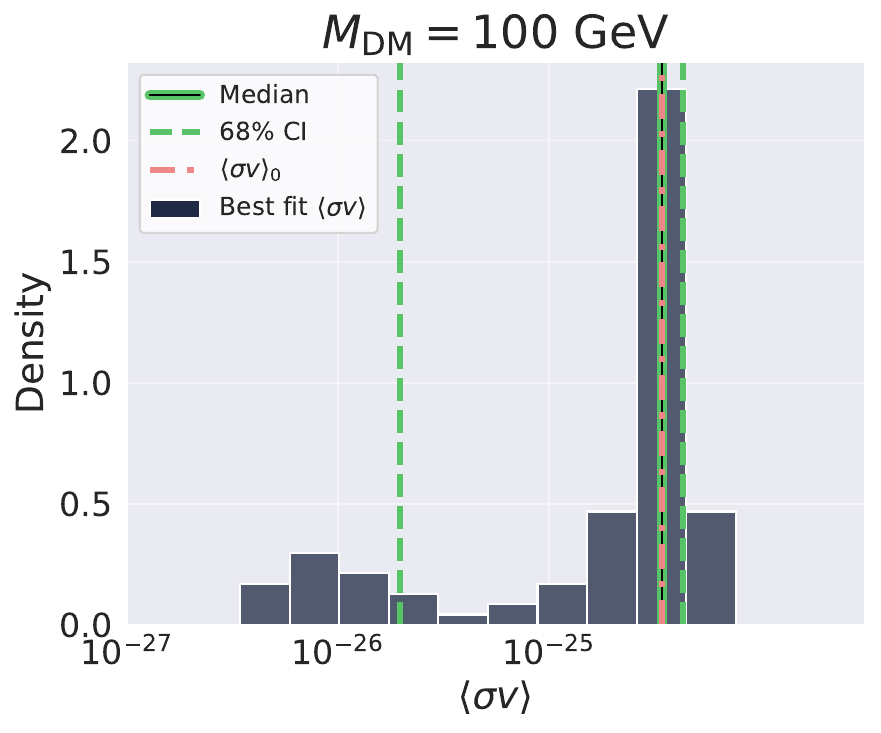}
    \includegraphics[width=0.45\linewidth]{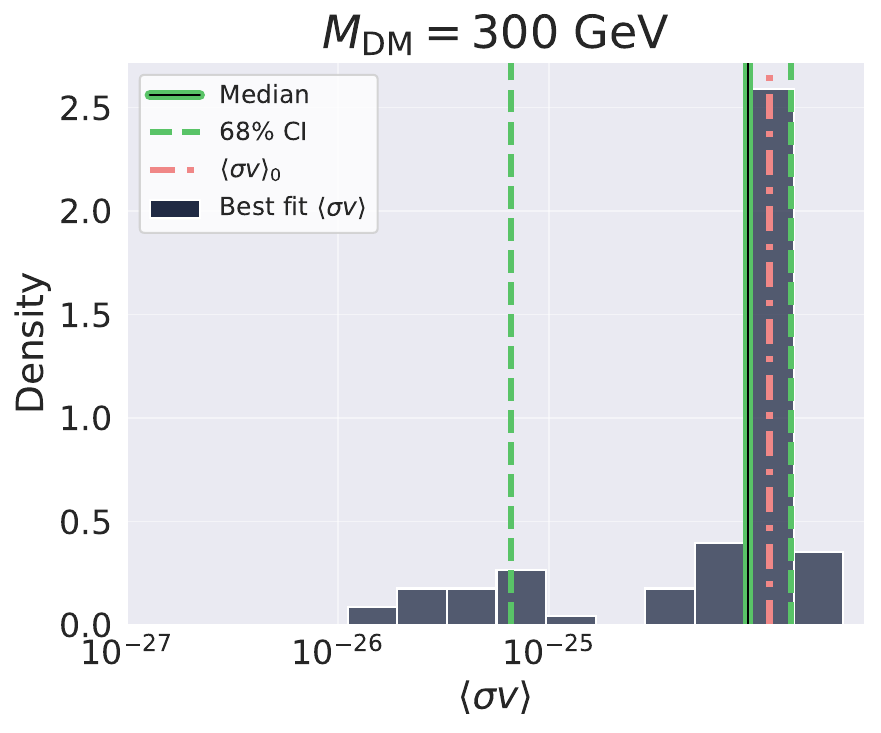}
    \includegraphics[width=0.45\linewidth]{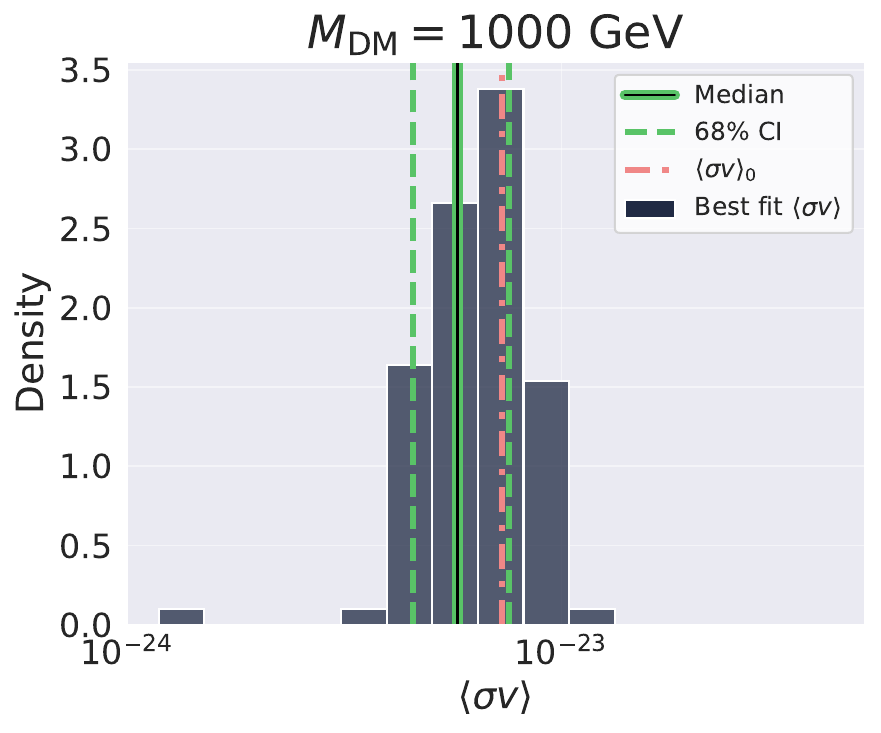}

    \caption{
    Histograms representing the reconstructed $\sigmav$ values for simulations with injected DM signals. The red dash-dotted line represents the annihilation cross-section of the injected signal ($\sigmav_0)$. The green solid line represents the median of the distribution, while the green dashed lines show the 68\% containment interval. In every case, the true $\sigmav$ value lies inside the 68\% containment region.
    }
    \label{fig:line-injection}
\end{figure}

We show the histogram of the recovered $\sigmav$ values for the 5 DM masses in figure~\ref{fig:line-injection}.
The $\sigmav$ values for the injected DM signal (red dashed-dotted lines, $\sigmav_0$) are within the 68\% containment interval (green dashed lines) of the histogram
relative to the median values for the distributions (solid green lines).
Thus, our methodology is capable of reconstructing a simulated signal, when present, in the majority of cases. 

\begin{figure}[t]
    \centering
    \resizebox{0.95\textwidth}{!}{
    \includegraphics[width=0.45\linewidth]{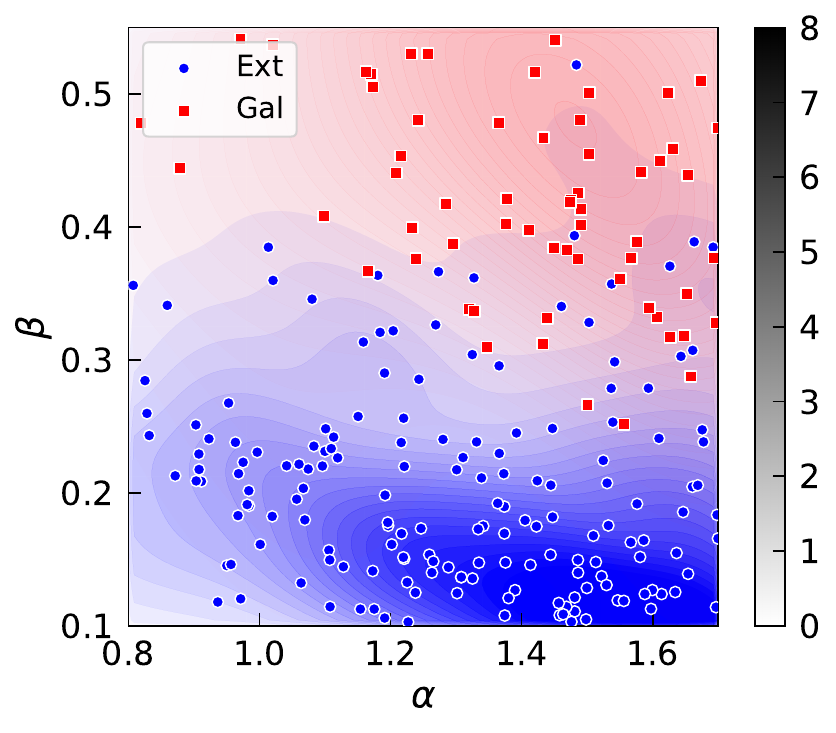}
    \includegraphics[width=0.45\linewidth]{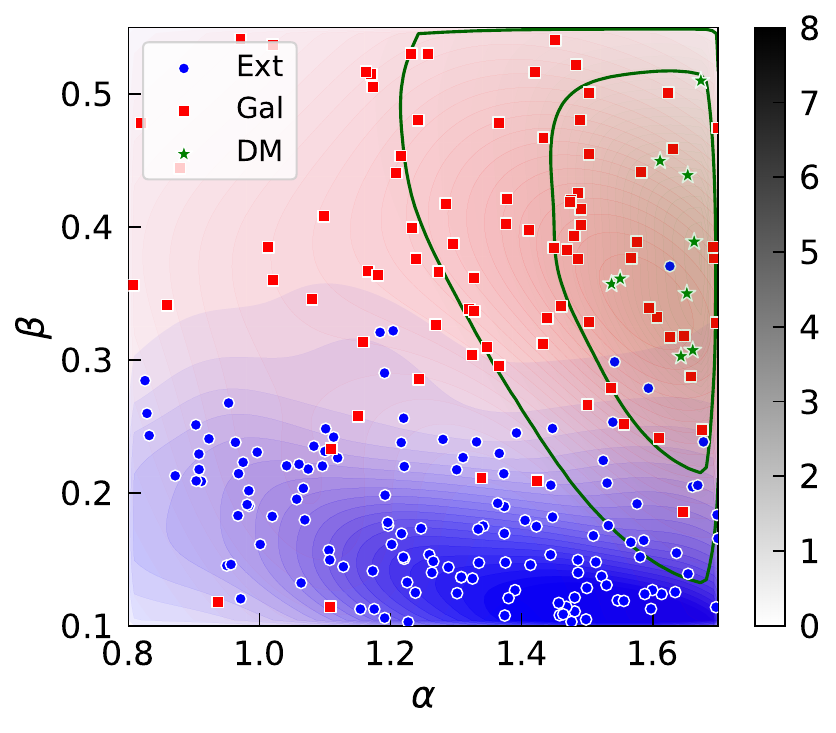}
    }
    \caption{Predicted source classes for unassociated 4FGL-DR4 sources according to the posterior class probabilities for the $\rm E_0 = 1 \,GeV$ case.
    \panel{Left panel}: best-fit model with astrophysical only classes.
    \panel{Right panel}: best-fit model for astrophysical classes and DM ($\mdm = 30$ GeV, $\sigmav \sim 2.7\cdot 10^{-25} \, {\rm cm}^{3}\,{\rm s}^{-1}$). The green contours represent the 68\% and 95\% containment levels for the DM distribution. 
    }
    \label{fig:predictions-real}
\end{figure}

Finally, we {illustrate} the model's predictive capabilities. Using eq.~(\ref{gamma}), we derive the posterior probability that each unassociated source belongs to a given class $k$ (e.g., extragalactic, Galactic, or DM). The predicted class label, $C_i^{\text{pred}}$, is then assigned to the class with the highest posterior probability: $C_i^{\text{pred}} = \operatorname{argmax}_k \{ \tilde{p}(z_{ik}=1|\mathbf{x}_i) \}$.
Figure~\ref{fig:predictions-real} displays the posterior probabilities and predicted classes for all unassociated sources. The left panel shows results from our best-fit model with only astrophysical classes (for $\mathrm{E}_0 = 1 \, \mathrm{GeV}$), while the right panel shows the best-fit model including DM (for the $\mdm=30$ GeV case).

\begin{figure}[t]
    \centering
    \resizebox{0.99\textwidth}{!}{
    \hspace{-3mm}
    \includegraphics[width=0.33\linewidth]{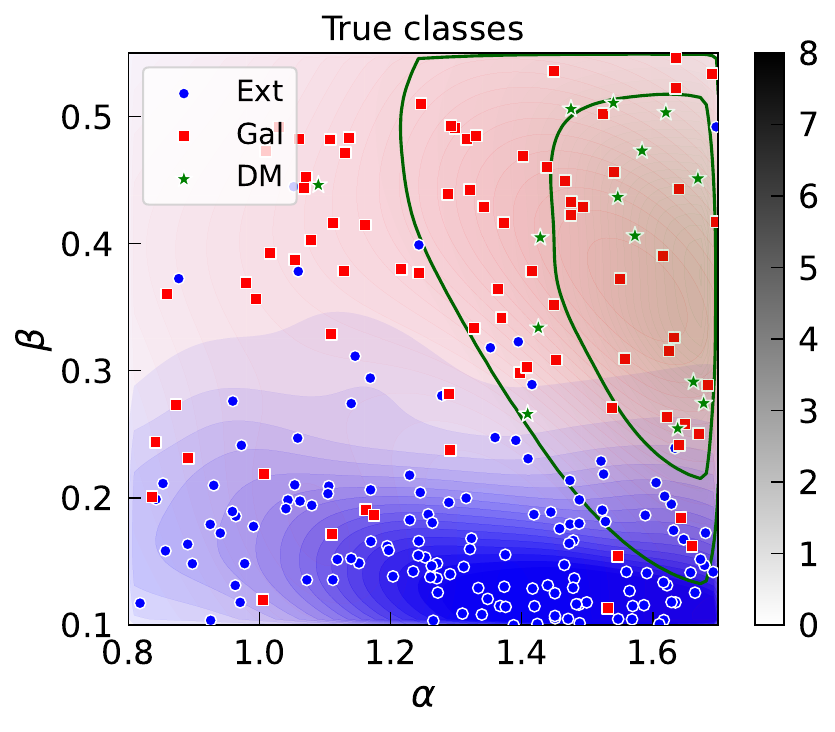}
    \hspace{-3mm}
    \includegraphics[width=0.33\linewidth]{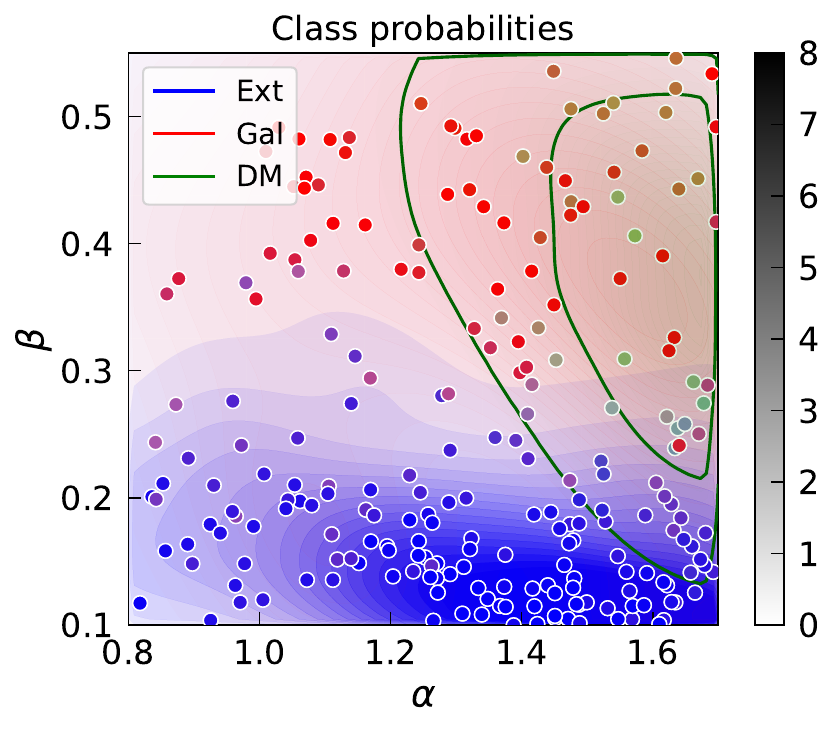}
    \hspace{-3mm}
    \includegraphics[width=0.33\linewidth]{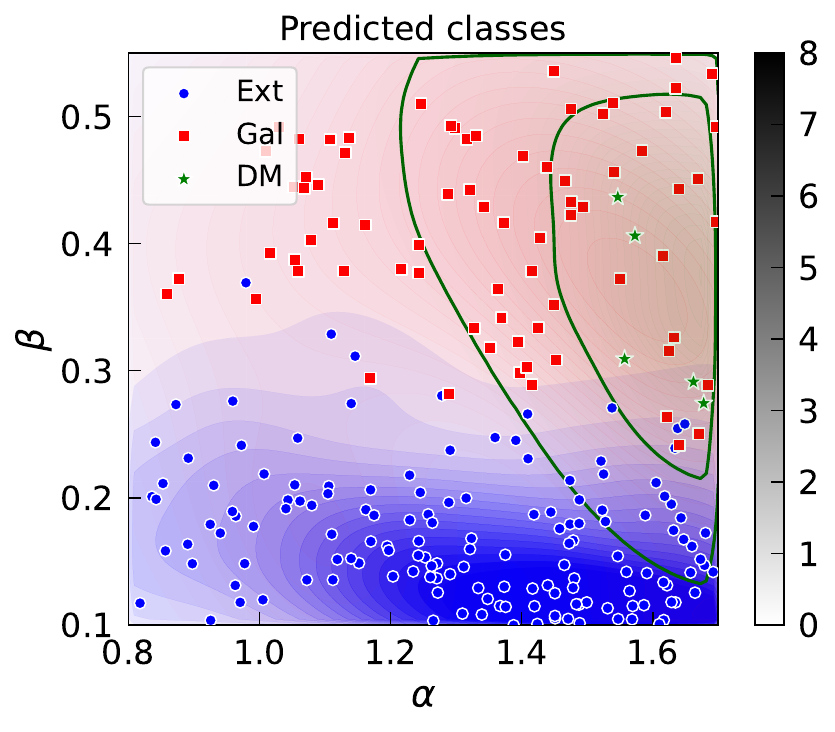}
    }
    \caption{Estimated best-fit distributions and true or predicted source classes for a bootstrap simulation, with $\mdm = 30$ GeV and $\sigmav \sim 2.7\times 10^{-25} \, {\rm cm}^{3}\,{\rm s}^{-1}$.
    \panel{Left panel}: markers colored according to their true class $C_i^{true}$ in the simulation.
    \panel{Middle panel}: markers colored according to their posterior class probability 
    following the same convention as figure~\ref{fig:mixture_DM_tot}.
    \panel{Right panel}: markers colored according to the predicted class, given by $C_i^{pred} = {\rm argmax}_k \{ \tilde{p}(z_{ik}=1|\mathbf{x}_i) \}$.
    }
    \label{fig:predictions-bootstrap}
\end{figure}

{For comparison,}
we repeat this exercise on a simulated bootstrap sample for the $\mdm=30$ GeV case in figure~\ref{fig:predictions-bootstrap}. 
The figure compares the true class labels (left panel) with the posterior class probabilities (middle panel) and the final predicted class labels (right panel) for a model fit to the bootstrap sample.
For instance, while not all DM sources were correctly classified due to significant parameter overlap with the Galactic source population, their true class often received a high posterior probability, even if it was not the maximum. This  {shows} that to avoid false negatives when identifying a subdominant class, it is {important} to consider sources with a {reasonably} high posterior probability for that class {even if this probability is not the highest}.

\subsection{{Precision-recall curves for classification of individual DM subhalos}}
\label{app:pr_curve}

{
To further quantify the model's ability to distinguish DM subhalos from the astrophysical background, we perform a Precision-Recall (PR) analysis for the case of $\mdm = 30$~GeV.
This DM mass yields the highest 
possible signal due to DM annihilation in subhalos (cf. table~\ref{tab:Nsub_values})
but it is also
challenging due to the extensive spectral overlap with Galactic sources. 
The PR framework is more relevant here than the Receiver Operating Characteristic curves for highly unbalanced datasets, as it focuses exclusively on the minority class (the signal), providing a rigorous assessment of how many ``DM-like" candidates are likely to be actual subhalos without being skewed by the overwhelming abundance of background sources.
}

\begin{figure}[t]
    \centering
    \resizebox{0.99\textwidth}{!}{
    \includegraphics[width=0.49\linewidth]{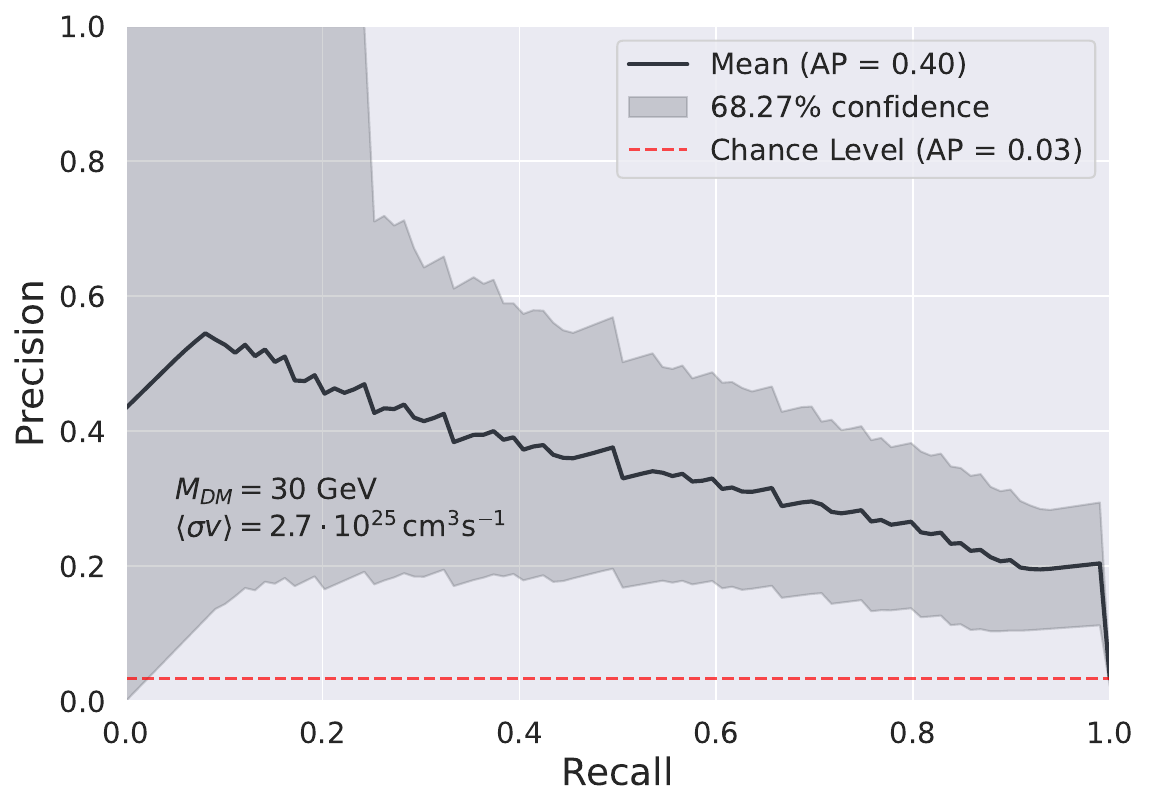}
    \includegraphics[width=0.49\linewidth]{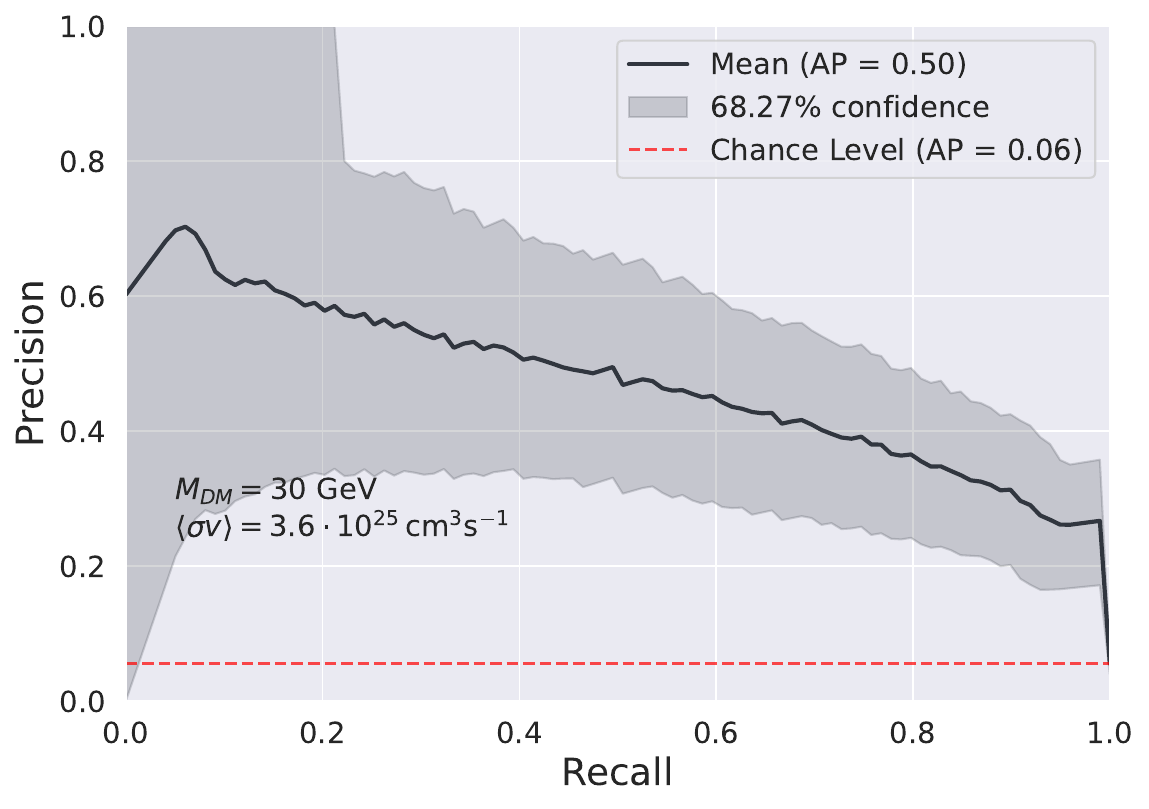}
    }
    \caption{
    {
    Precision and recall for classification of individual DM subhalos among unassociated \Fermi-LAT sources.
    Solid black line (gray band) shows the mean (68.27\% statistical containment) for 1000 simulations at $\mdm = 30$~GeV. \panel{Left panel}: performance for the best-fit cross-section in table~\ref{tab:Nsub_values} (average precision $AP = 0.40 \pm 0.16$). \panel{Right panel}: performance for a cross-section close to the 95\% confidence upper bound in table~\ref{tab:Nsub_values} ($AP = 0.50 \pm 0.14$).}
    }
    \label{fig:pr-curves}
\end{figure}

{
We consider two physical scenarios for the annihilation cross-section: the best-fit value obtained in our primary analysis ($\sigmav = 2.7 \times 10^{-25} \text{ cm}^3/\text{s}$) and a value close to the 95\% upper bound ($\sigmav = 3.6 \times 10^{-25} \text{ cm}^3/\text{s}$). For each scenario, we evaluate the PR performance across a suite of 1000 independent simulations.
In figure~\ref{fig:pr-curves}, we display the mean PR curve along with the 68.27\% asymmetric confidence band derived from the simulation quantiles, representing the expected statistical variance due to background realizations and Poisson noise.
}
\\
{
To quantify performance, we compute the Average Precision (AP):
\be
AP = \sum_n (R_n - R_{n-1}) P_n
\ee
where $P_n$ and $R_n$ are the precision and recall at the $n$-th classification threshold. We find $AP = 0.40 \pm 0.16$ for the best-fit case and $AP = 0.50 \pm 0.14$ for the upper bound. 
These scores significantly outperform their respective random chance baselines of 0.03 and 0.06.
The fact that the AP remains below unity is a direct reflection of the intrinsic physical confusion between DM subhalos annihilating in the $b\bar{b}$ channel for $\mdm = 30$~GeV and Galactic millisecond pulsars.
}

\end{document}